\documentclass[journal=jpcafh,manuscript=article]{achemso}
\usepackage{longtable}
\usepackage{multirow}
\usepackage{amsmath}
\usepackage{subfigure}
\usepackage[usenames,dvipsnames]{xcolor}
\usepackage{soul}
\usepackage{bm}

\usepackage{xr}
\usepackage{amssymb}
\usepackage{siunitx}
\usepackage{threeparttable}
\usepackage{multicol} \usepackage{wrapfig} \usepackage{enumitem}
\usepackage{bm} \usepackage{outlines} %for itemize with subitems
\usepackage{booktabs}
\usepackage{chemfig}

\makeatletter
\newcommand*{\addFileDependency}[1]{% argument=file name and extension
  \typeout{(#1)}
  \@addtofilelist{#1}
  \IfFileExists{#1}{}{\typeout{No file #1.}}
}
\makeatother

\newcommand*{\myexternaldocument}[1]{%
    \externaldocument{#1}%
    \addFileDependency{#1.tex}%
    \addFileDependency{#1.aux}%
}
\myexternaldocument{si}
\SectionNumbersOn
\author{Silvan K\"aser} \affiliation[University of Basel]{Department
  of Chemistry, University of Basel, Klingelbergstrasse 80, CH-4056
  Basel, Switzerland.}

\author{Debasish Koner}\affiliation[IISER Tirupati]{Department of
  Chemistry, Indian Institute of Technology Hyderabad, Kandi,
  Sangareddy-502285, Telangana,
  India}

\author{Markus Meuwly} \affiliation[University of Basel]{Department of
  Chemistry, University of Basel, Klingelbergstrasse 80, CH-4056
  Basel, Switzerland.}\email{m.meuwly@unibas.ch}

\title{The Bigger the Better? Accurate Molecular Potential Energy
  Surfaces from Minimalist Neural Networks}

\begin{document}
\date{\today}
%TC:ignore
\begin{abstract}
Atomistic simulations are a powerful tool for studying the dynamics of molecules, proteins, and materials on wide time and length scales. Their reliability and predictiveness, however, depend directly on the accuracy of the underlying potential energy surface (PES). Guided by the principle of parsimony this work introduces KerNN, a combined kernel/neural network-based approach to represent molecular PESs. Compared to state-of-the-art neural network PESs the number of learnable parameters of KerNN is significantly reduced. This speeds up training and evaluation times by several orders of magnitude while retaining high prediction accuracy. Importantly, using kernels as the features also improves the extrapolation capabilities of KerNN far beyond the coverage provided by the training data which solves a general problem of NN-based PESs. KerNN applied to spectroscopy and reaction dynamics shows excellent performance on test set statistics and observables including vibrational bands computed from classical and quantum simulations.
\end{abstract}
%TC:endignore

%\section{Introduction}
Molecular dynamics (MD) simulations have long been a cornerstone of
research in fields ranging from chemistry and biology to materials
science and drug
discovery.\cite{van1990computer,karplus1990molecular,durrant2011molecular}
Such simulations offer a unique lens through which the dynamic
behaviour of molecules and materials can be explored at the atomic
scale, enabling the study of molecular interactions and complex
physical and chemical processes. Central to the success of MD
simulations is the availability of an energy function such as an
empirical force field (FF) or more generally a potential energy
surface (PES) to obtain the total energy of the system. In the case of
FFs the function relies on empirically chosen mathematical functions and
parameters, such as the popular CHARMM general force field (CGenFF)
for drug-like
molecules.\cite{vanommeslaeghe2010charmm,hwang2024charmm}\\

\noindent
While traditional FFs have played a crucial role in advancing our
understanding of molecular systems, they are not without
limitations. One of the most notable challenges lies in achieving a
balance between accuracy and computational efficiency. This trade-off
between accuracy and speed has been a persistent bottleneck in the
field of atomistic simulations. Recent advances in machine learning
(ML), particularly in the field of kernel-based and neural network
(NN) methods, have revolutionised our ability to model molecular
systems with unprecedented
precision.\cite{unke2021machine,kaser2023neural,qu2018assessing} Such
machine learned PESs (ML-PESs) can be trained to reproduce the quality
of the underlying quantum chemical reference data with remarkable
accuracy. This also eliminates the need for extensive and
time-consuming empirical parametrization.\\

\noindent
Although ML-PESs reach extraordinary accuracies with respect to their
\textit{ab initio} reference, their computational cost still lies
between \textit{ab initio} techniques and traditional empirical FFs
which compromises their applicability in long-time and large-scale MD
simulations.\cite{unke2021machine,wang2024design} Since the accurate
computation of experimental observables - such as optical spectra,
pair distribution functions, free energies, to name a few - requires
sufficiently long and accurate MD simulations, even ML-based PESs
eventually become impractical due to their slow evaluation speed. In
addition, typical NN-based approaches suffer from poor extrapolation
properties, in particular in regions not covered by training data. To
address these issues the present work introduces small NN-based models
for molecular PESs which speeds up training and in particular
evaluation times by $\sim 2$ orders of magnitude for the systems
considered. This constitutes a significant step forward in view of
larger-scale, accurate, and resource-efficient MD simulations using
ML-based energy functions.\\

\begin{figure}[ht]
\centering
\includegraphics[width=0.9\textwidth]{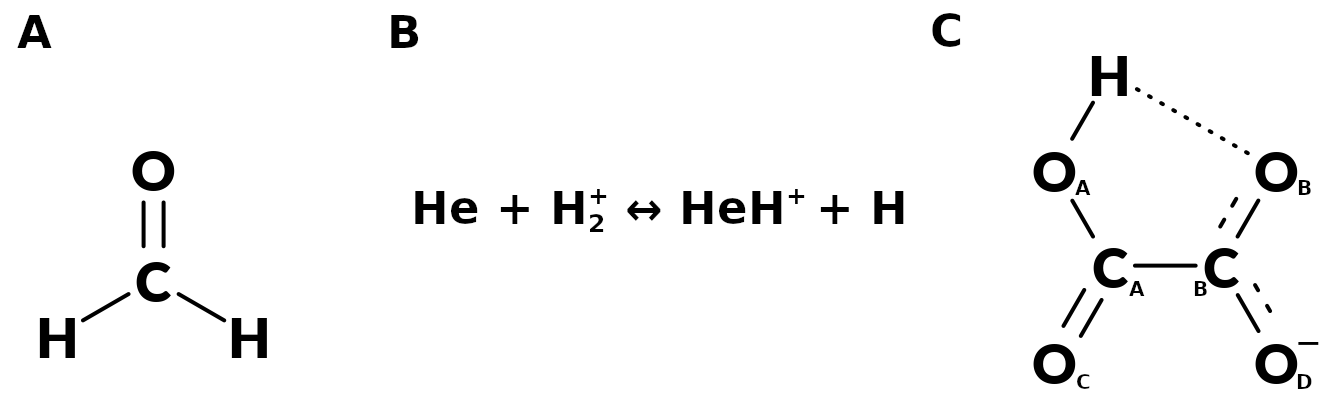}
\caption{Schematic representation of A) formaldehyde, B) the two
  reaction channels of the HeH$_2^+$ system and C) hydrogen oxalate
  (or deprotonated oxalic acid) in its cyclic/hydrogen bonded form.}
\label{fig:schematics_molecules}
\end{figure}

\noindent
To increase the computational efficiency in training and evaluating
ML-PESs the present work proposes to combine reproducing kernel Hilbert space (RKHS)
techniques with
NN-based strategies which will be referred to as KerNN in the
following. First, the approach is applied to H$_2$CO as a benchmark
system. Next, KerNN is applied to two reactive systems including
HeH$_2^+$ and the hydrogen oxalate anion (or deprotonated oxalic acid,
HO$_2$CCO$_2^-$). Schematics of the three molecular systems are given
in Figure~\ref{fig:schematics_molecules}. Finally, the performance of
KerNN in terms of accuracy, efficiency and robustness of the resulting
simulations are presented and discussed in a broader context.\\

\section*{Results and Discussion}
\subsection*{KerNN Architecture}
KerNN is based on a small feed-forward NN with one input, two hidden
and one output layer. The molecular descriptors are one-dimensional
reciprocal power reproducing kernels, which effectively serve as a
similarity measure between the interatomic distances of a
reference and a query structure. Both,
non-permutationally invariant ($\mathcal{D}^{\rm ns}$) and
permutationally invariant ($\mathcal{D}^{\rm s}$) descriptors
constructed from fundamental invariants\cite{derksen2015computational} 
are employed. The total
potential energy of the system is KerNN's output and forces are
calculated via reverse mode automatic differentiation.\cite{baydin2017automatic}
For spectroscopic applications, KerNN was adapted to
predict dipole moments, too. Details are provided in the Methods
section.\\

\subsection*{H$_2$CO}
\begin{figure}[ht]
\centering
\includegraphics[width=1.0\textwidth]{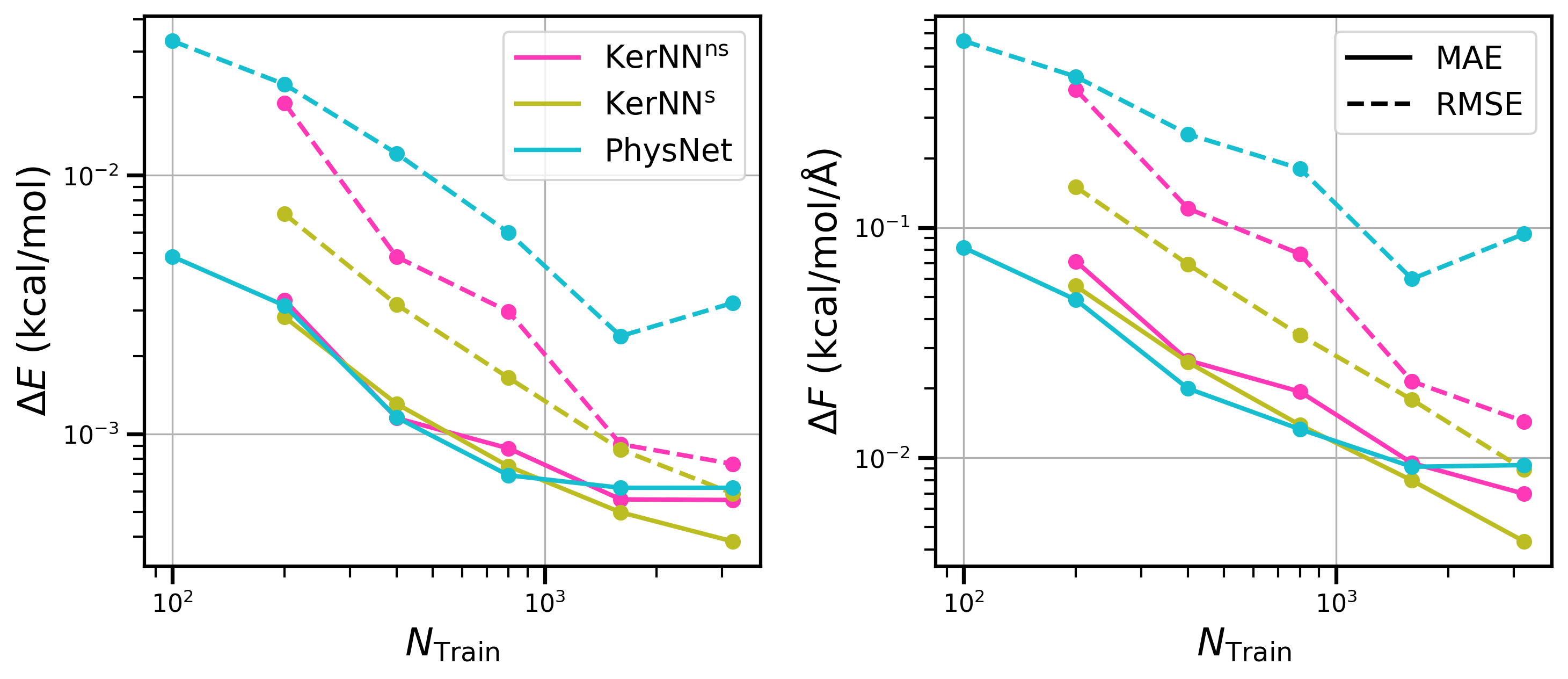}
\caption{Energy and force learning curves for the different variants
  of the H$_2$CO PESs trained on CCSD(T)-F12B/aug-cc-pVTZ-F12
  reference data. These are compared to PhysNet results taken from
  Reference~\citenum{MM.h2co:2020}. Solid and dashed lines represent
  MAEs and RMSEs, respectively. A total of five KerNN models were
  trained for each value of $N_{\rm Train}$ on different splits of the
  data and only the mean out-of-sample errors are shown. KerNN$^{\rm
    ns}$ and KerNN$^{\rm s}$ represent the NNs that use the
  non-symmetrized ($\mathcal{D}^{\rm ns}$) and symmetrized
  ($\mathcal{D}^{\rm s}$) descriptors, respectively. The flattening in
  energies for $N_{\rm Train} \geq 1600$ is caused by the "error
  floor" noted in earlier work for the CCSD(T)-F12
  forces.\cite{MM.h2co:2020} Note that the different models can be
  more or less sensitive to such noise and therefore exhibit a
  flattening at higher/lower test set errors. The lowest test set MAEs
  reported in Reference~\cite{MM.h2co:2020}, for example, were
  MAE($E$) = 3E-4 and MAE($F$)=1E-4.}
\label{fig:kernn_lc}
\end{figure}

Training of KerNN for H$_2$CO was based on non-symmetrized
$\mathcal{D}^{\rm ns}$ and symmetrized $\mathcal{D}^{\rm s}$
descriptors. Five independent repeats of training on different splits
of the data and using four different data set sizes $N_{\rm Train} =
[400, 800, 1600, 3200]$ were carried out. The validation set
invariably consisted of $N_{\rm Valid} = 400$ structures while the
hold-out test set contained the remaining $N_{\rm Test} = N_{\rm Tot}
- (N_{\rm Train} +N_{\rm Valid})$ structures. Energy and force
learning curves for KerNN$^{\rm ns}$ (NN employing the non-symmetrized
descriptor $\mathcal{D}^{\rm ns}$, given in
Equation~\ref{eq:descriptor_h2co}) and KerNN$^{\rm s}$ (NN employing
the symmetrized descriptor $\mathcal{D}^{\rm s}$, given in
Equation~\ref{eq:descriptor_h2co}) are shown in
Figure~\ref{fig:kernn_lc}.\\

\noindent
Learning curves quantify the rate at which ML models learn. It has
been shown empirically\cite{muller1996numerical} that on average the
test set errors must decay inversely with training set size $N_{\rm
  Train}$ according to a power law
\begin{align}
    {\rm Error} \approx  a/N_{\rm Train}^b.
\end{align}
On a log-log plot, learning curves for ML models should
therefore follow
\begin{align}
    \log({\rm Error}) \approx  \log(a) - b\log(N_{\rm Train})
\end{align}
with an offset $\log(a)$ and a slope
$b$.\cite{huang2016communication,christensen2020role} The learning
curves in Figure~\ref{fig:kernn_lc} demonstrate that the logarithm of
the error consistently decreases linearly for larger training set
sizes for both, $\mathcal{D}^{\rm ns}$ and $\mathcal{D}^{\rm s}$. This
is particularly evident for the force learning curves. The quantity
that was minimized during training was a mean squared error loss, in
which the errors in the forces contributed ten times more ($\omega_F =
10$, see Table~\ref{sitab:nn_hyperparams}).  Thus, the most meaningful
comparison of the different models in Figure~\ref{fig:kernn_lc} is
given by RMSE($F$) (right panel, dashed lines). Here, KerNN$^{\rm s}$
reaches the lowest out-of-sample errors throughout. Notably, the
approach presented herein can also be compared to other
state-of-the-art ML models in terms of out-of-sample errors on the
same data set.\cite{MM.h2co:2020} With $N_{\rm Train} = 200$ the
MAE($F$) (RMSE($F$)) for the two KerNN variants are $7.1\cdot10^{-2},
5.5\cdot10^{-2}$ (0.40, 0.15)~kcal/mol/\AA~, compared with
$4.9\cdot10^{-2}$, $6.1\cdot10^{-2}$ and $5.1\cdot10^{-4}$ (0.45, 0.33
and $3.0\cdot10^{-3}$)~kcal/mol/\AA\/ for
PhysNet\cite{MM.physnet:2019}, RKHS+F\cite{MM.rkhs:2020} and kernel
ridge regression using the FCHL descriptor\cite{faber2018alchemical},
respectively.\\

\begin{figure}[h!]
\centering
\includegraphics[width=0.8\textwidth]{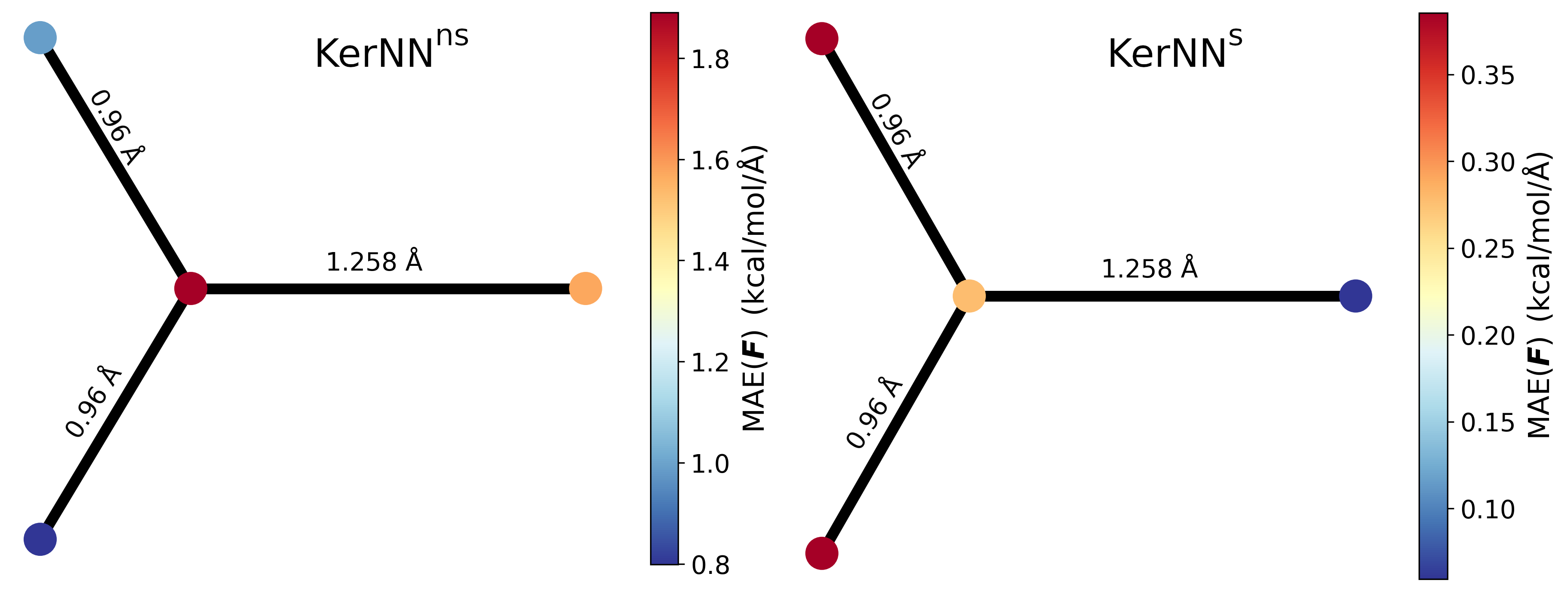}
\caption{Atomic MAE($\bm{F}$) (\textit{i.e.} the mean absolute error
  between the reference and predicted forces for each of the atoms)
  for the symmetric structure with short H-C bonds shown. The
  corresponding (aggregated) MAE($\bm{F}$) are
  1.3/0.3~kcal/mol/\AA. While KerNN$^{\rm s}$ predicts symmetrical
  errors, this is not so for KerNN$^{\rm ns}$.}
\label{fig:atomic_mae}
\end{figure}

\noindent
It is interesting to assess to what extent permutational invariance of
the identical H-atoms impacts the predictions ($E$ and $\bm{F}$) of
KerNN$^{\rm ns}$ and KerNN$^{\rm s}$. For KerNN$^{\rm s}$ the total
energy is invariant upon exchange of the two H-atoms by construction,
which is not guaranteed for KerNN$^{\rm ns}$ (or any other approach
that is not permutationally invariant). However, for the optimized
H$_2$CO configuration the energy of KerNN$^{\rm ns}$ remains unaltered
upon H-exchange. This is a consequence of the symmetry in the data and
also because the small size of H$_2$CO helps to cover the relevant
symmetry-related structures. A similar analysis can be carried out for
forces by predicting them for a symmetric (C$_{\rm 2v}$) structure of
H$_2$CO and comparing them to their \textit{ab initio}
counterparts. The analysis of the atomic MAE($\bm{F}$), shown in
Figure~\ref{fig:atomic_mae}, illustrates that the atomic MAE($\bm{F}$)
are larger for the structure using KerNN$^{\rm ns}$. Also, KerNN$^{\rm
  ns}$ fails to predict fully symmetric forces (with atomic
MAE($\bm{F}$) of 0.99 and 0.80~kcal/mol for the two H-atoms) whereas
for KerNN$^{\rm s}$ the atomic MAE($\bm{F}$) is identical on both
H-atoms (0.39~kcal/mol).\\

\noindent
An additional test for the accuracy of the forces (or the curvature of
the PES) around a stationary point are the harmonic frequencies. These
are given in Table~\ref{sitab:kernn_harmfreq} and compared to their
\textit{ab initio} CCSD(T)-F12B reference and frequencies obtained
from other ML approaches,\cite{MM.h2co:2020} including
PhysNet\cite{MM.physnet:2019}, RKHS+F\cite{MM.rkhs:2020} and kernel
ridge regression using the FCHL descriptor.\cite{faber2018alchemical}
KerNN$^{\rm ns}$ features excellent accuracy
(MAE($\omega$)=0.2~cm$^{-1}$) and is competitive in comparison to
both, KerNN$^{\rm s}$ and the more sophisticated (in terms of NN
architecture, physically motivated terms such as electrostatics,
number of parameters, descriptors that include distance and angular
terms, \textit{etc.})  ML approaches (MAE($\omega$)=0.1~cm$^{-1}$).\\

\begin{figure}[h!]
\centering
\includegraphics[width=0.85\textwidth]{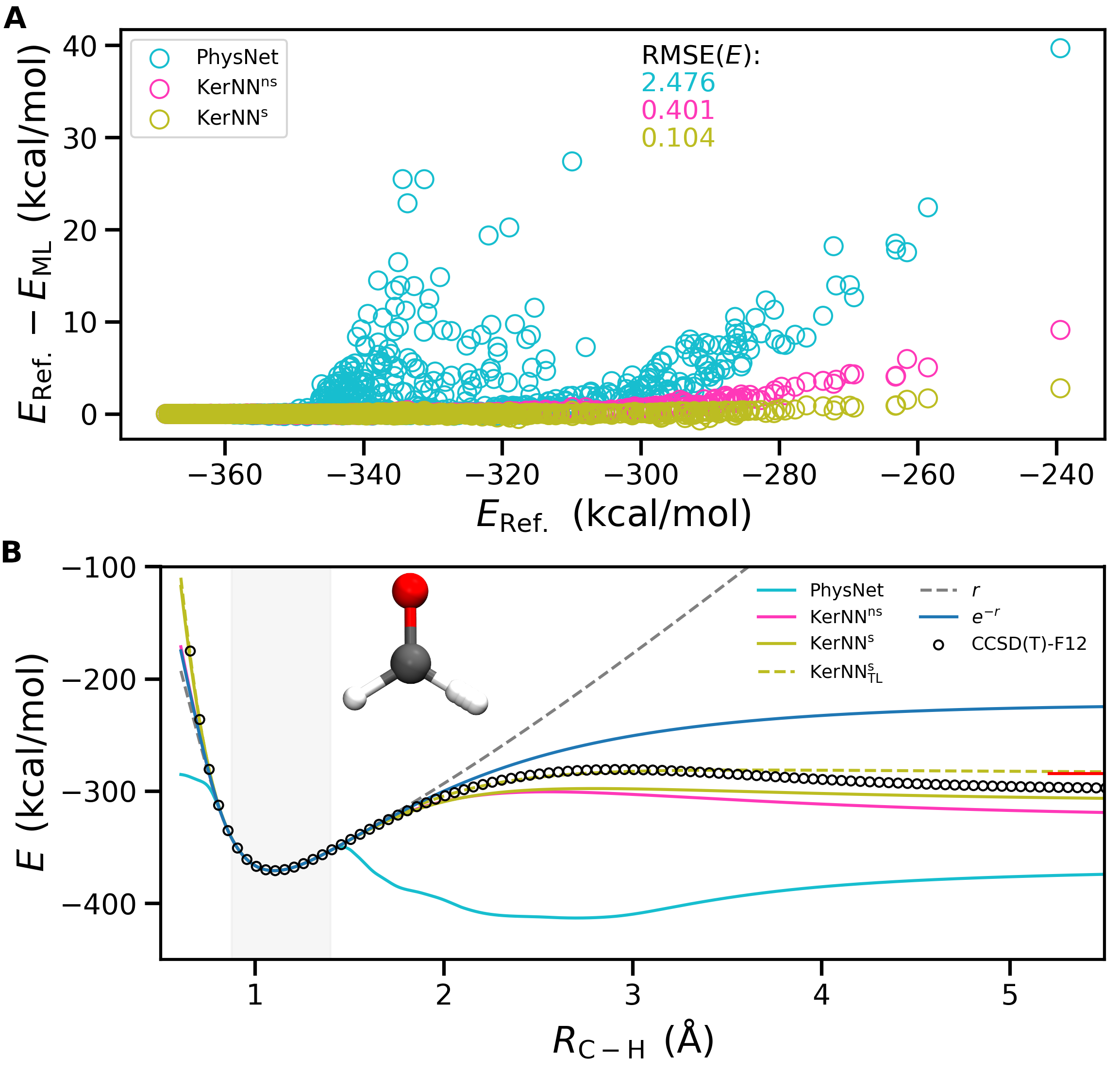}
\caption{A: The extrapolation capabilities of the ML-PES are assessed
  on a data set containing 2500 structures, generated from normal mode
  sampling at a higher temperature (5000~K) than the training set
  ($2000$~K). The extrapolation data set was available from previous
  work.\cite{MM.h2co:2020} While the training data covers an energy
  range of roughly 40 kcal/mol the extrapolation data set covers
  130~kcal/mol. B: One-dimensional PES cut along the C-H bond length
  for different ML models ($r$ and $e^{-r}$ correspond to NNs with the
  same architecture as KerNN, but employing different descriptors,
  namely the interatomic distances $r$ and $e^{-r}$). KerNN$^{\rm
    s}_{\rm TL}$ corresponds to a model with asymptotic behaviour
  adjusted according to the experimentally determined dissociation
  energy. The training data range is shaded in gray.}
\label{fig:extrapolation_h2co}
\end{figure}

\noindent
The previous evaluations focused on analyzing the models' capability
for interpolation, \textit{i.e.}, how well they predict the properties
of structures within (\textit{e.g.}, energy- or interatomic
distance-wise) their training data.  A more challenging task, however,
concerns the extrapolation capabilities of ML
models. Reliable extrapolation is not at all guaranteed because such
models are purely mathematical constructs without inherent physical
meaning. Specifically for PESs to be used for dynamics simulations,
poor extrapolation capabilities can lead to significant errors,
potentially resulting in unphysical or incorrect behavior in areas of
the energy landscape that are not covered by the training data. As was
noted in References~\citenum{tea2024analysis} and \citenum{tea2024md},
low test set errors do not guarantee robust MD simulations. Hence, the
extrapolation performance of both KerNN variants was considered, see
Figure~\ref{fig:extrapolation_h2co}. The extrapolation dataset, shown
in Figure~\ref{fig:extrapolation_h2co}A and available from previous
work\cite{MM.h2co:2020}, covers structures that were sampled at
considerably higher temperatures (5000 K) than the training set
(2000~K) and covers a much broader energy range (130
vs. 40~kcal/mol). Notably, KerNN$^{\rm s}$ extrapolates very
accurately without any extreme outliers and has only small deviations
at the highest energies. KerNN$^{\rm ns}$ closely follows the accuracy
of KerNN$^{\rm s}$ with slightly larger deviations in the high energy
range.\\

\noindent
Meaningful extrapolation outside the range covered by the
training/validation/test data is typically very challenging for ML
models for both, NNs and kernel-based methods (depending on what
descriptor is used). Both tend to behave unphysically and have
increasingly large prediction errors in the extrapolation
regime.\cite{unke2021machine} An example is reported in
Figure~\ref{fig:extrapolation_h2co}B which shows one-dimensional cuts
through different PESs along a C-H bond of formaldehyde. The
evaluation illustrates the robust extrapolation of both KerNN variants
(magenta, olive) far beyond the training data whereas PhysNet (cyan)
fails for bond lengths outside the training data set.  As a
comparison, two additional descriptors were employed for the same NN
architecture as KerNN (non-symmetrized set of interatomic distances
(r, gray-dashed) and non-symmetrized exponentials of the set of
negative interatomic distances ($e^{-r}$, blue)). It is conjectured
that the long-range asymptotics of KerNN can be further adjusted and
controlled by including targeted data from either {\it ab initio}
calculations or experiments.  This is valuable for situations in which
electronic structure calculations fail to provide reliable reference
data as can be the case for MRCI or CASPT2 calculations in the
dissociation region.\\

\noindent
To test this, the experimentally determined C-H dissociation energy
(86.6 kcal/mol) was used to further constrain the long range part of
the KerNN PES.\cite{chuang1987t1} Ten structures with a C-H distance
of $\sim 10$ \AA\/ were assigned with that energy and the model was
transfer learned by fine-tuning the weights and biases of the
KerNN$^{\rm s}$ PES. The energy of the empirical data points was set
to (-371 + 86.6 = ) -284~kcal/mol and is marked by a red line in
Figure~\ref{fig:extrapolation_h2co}, to which the KerNN$^{\rm s}_{\rm
  TL}$ is successfully corrected. It is noted that the \textit{ab initio} reference
data predicts a small barrier at $R_{\rm C-H} \sim 3$~\AA\/ and it is
unclear if this a spurious barrier caused by using a single-reference
method for a multi-reference problem. Such spurious barriers (or
minima) can have detrimental effects on the dynamics of a molecular
system.\cite{dawes2016single} The reliable extrapolation capability of
KerNN is likely a consequence of using kernels as descriptors since
$k(r_i, r_i')$ decays smoothly and monotonically towards zero for
large $r_i$ (see Figure~\ref{sifig:kernn_heh2p_1dk}). In other words,
the guaranteed long-range decay of $k(r_i, r_i')$ as a feature in the
NN controls the global behaviour of the energetics and avoids
arbitrary predictions outside the range covered by the training
data. This is an essential advantage of KerNN over other NN-based and
in part other kernel-based approaches.\\

\noindent
Finally, KerNN can also be used to run finite-temperature MD
simulations from which a multitude of experimental observables can be
computed. Here, the infrared spectrum (IR) was determined which was
also available from earlier studies of H$_2$CO using
PhysNet\cite{MM.h2co:2020} and from experiments.\cite{herndon:2005}
For that reason, KerNN$^{\rm ns}$ was trained on energies, forces and
dipole moments and its test set errors are given in
Table~\ref{sitab:kernn_h2co_efmu}. IR spectra were determined from the
Fourier transform of the dipole moment autocorrelation function
\cite{topfer2023molecular} following
\begin{equation}
  I(\omega) n(\omega) \propto Q(\omega) \cdot \mathrm{Im}\int_0^\infty
  dt\, e^{i\omega t} 
  \sum_{i=x,y,z} \left \langle \boldsymbol{\mu}_{i}(t)
  \cdot {\boldsymbol{\mu}_{i}}(0) \right \rangle.
\label{eq:IR}
\end{equation}
The Fourier transform was multiplied by a quantum correction factor
$Q(\omega) = \tanh{(\beta \hbar \omega / 2)}$.\cite{kumar2004quantum}
The dipole moment was that from the trained KerNN$^{\rm ns}$ model
which incorporates effects due to atom-centered fluctuating
charges. Unfortunately, learning the dipole moment by training on the
extended loss function (\ref{eq:loss_kernn_dipole}) is impossible for
KerNN$^{\rm s}$ due to its permutational invariance (see
Figure~\ref{sifig:kernns_h2co_dipole_inability}).\\

\begin{figure}[ht]
\centering
\includegraphics[width=1.0\textwidth]{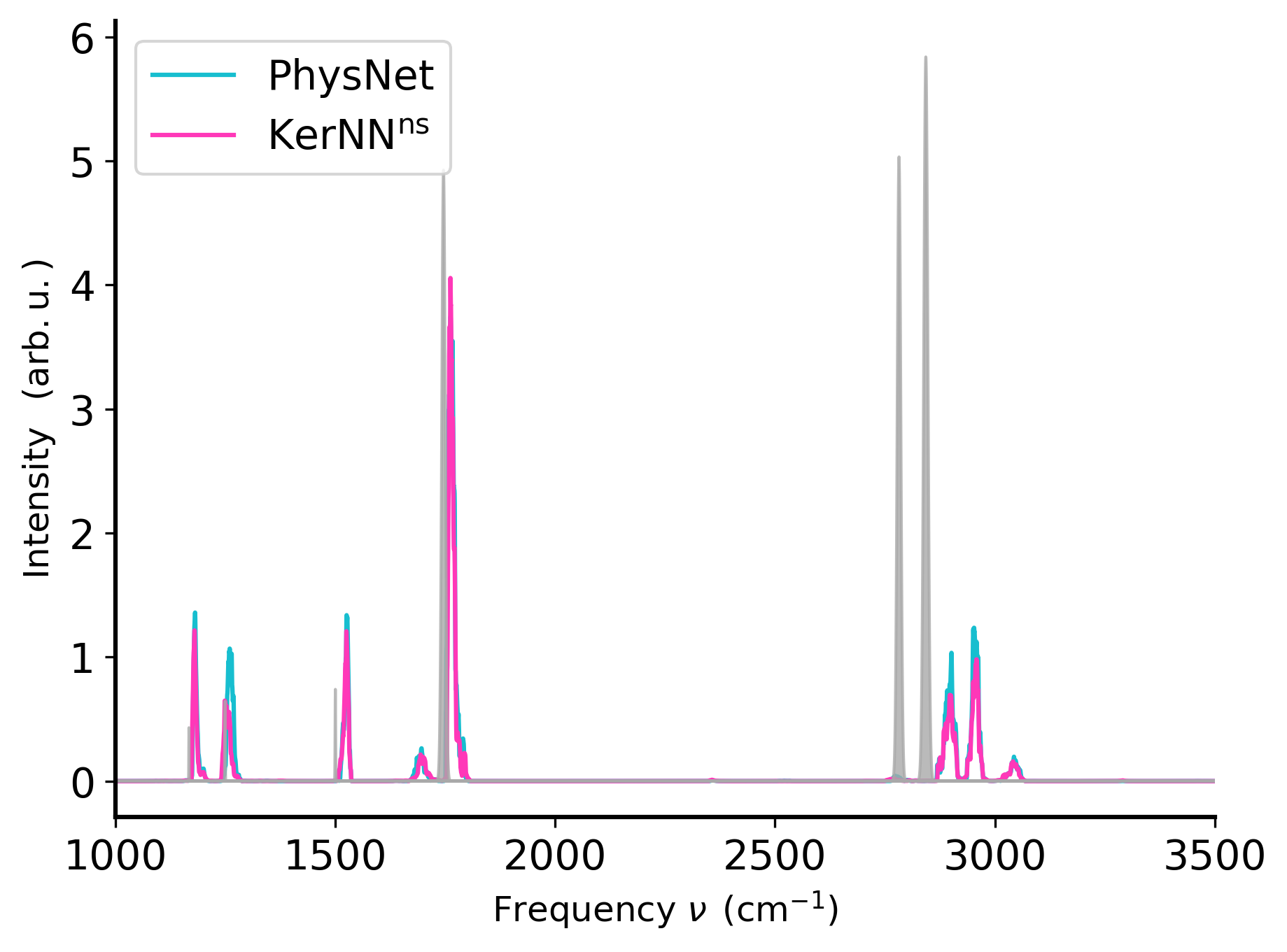}
\caption{Infrared spectra derived from finite-$T$ MD simulations of
  H$_2$CO. The experimental fundamentals\cite{herndon:2005} are the
  grey Gaussians. The computed spectra were averaged over 100
  independent trajectories, each 200~ps in length, using $\Delta t =
  0.2$~fs.}
\label{fig:kernn_h2co_ir_comp}
\end{figure}

\noindent
The MD simulations, carried out using the atomic simulation environment (ASE)\cite{larsen2017atomic}, were each 200 ps in length,
run with a time step of $\Delta t = 0.2$ fs to conserve total energy,
starting from randomly initialized momenta drawn from a
Maxwell-Boltzmann distribution at $T = 500$ K and using the optimized
H$_2$CO structure as the initial configuration. Averaged IR spectra
determined from 100 independent simulations are shown in Figure
\ref{fig:kernn_h2co_ir_comp}. In terms of peak positions, the
KerNN$^{\rm ns}$ and PhysNet spectra are indistinguishable, and only
small differences in their intensities are found. In addition to the
six fundamental bands, an overtone and a combination band are visible
at $\sim 2770$ and 3040~cm$^{-1}$, respectively. Comparable to the
computed IR spectrum, the experimental spectrum of pure formaldehyde
ice\cite{jolly:2015} also shows a sideband on the red side of the 1725
cm$^{-1}$ peak which, however, remained unassigned.\\

\noindent
From a more technical perspective, the results so far are particularly
remarkable and promising considering the small number of learnable
parameters (1001 and 1021 for KerNN$^{\rm ns}$ and KerNN$^{\rm s}$)
the two KerNN approaches have. This compares with kernel-based methods
that usually have one parameter per data point and NNs that can have
millions of parameters. The compact form and resulting inference speed
of KerNN is the advantage of such approaches and their computational
cost for energy and force evaluations is given in
Table~\ref{sitab:kernn_timings}. KerNN in its Python implementation is
roughly 15 (50) times faster than PhysNet (kernel ridge regression
with FCHL), and only slightly slower than RKHS+F, which is written in
FORTRAN.\cite{MM.rkhs:2017,MM.rkhs:2020} If KerNN's FORTRAN
implementation is used (\textit{e.g.}, from within CHARMM), this
yields a 100-fold speedup over its Python implementation and,
therefore, outperforms PhysNet by orders of magnitude. Compared with
RKHS+F, the compact NNs are 20 to 30 times faster allowing
considerably longer simulation times at comparable accuracy. Notably,
if the training set size is increased (as would likely be necessary
for larger molecules), the computational cost of an $E$ and $\bm{F}$
evaluation remains the same for KerNN, whereas it scales quadratically
with the training/reference data for RKHS+F and the matrix inversion
for obtaining the linear coefficients scales cubically which becomes
prohibitive for reference data larger than $\sim 10^4$.\\

\subsection*{HeH$_2^+$}
Next, KerNN was applied to a reactive system: ${\rm He-H}_2^+
\rightarrow {\rm HeH}^+ + {\rm H}$. Breaking and forming chemical
bonds requires a {\it globally valid} PES, as opposed to a local PES
as was the case for H$_2$CO, which is generally more
challenging. Again, two descriptors, $\mathcal{D}^{\rm ns}$ and
$\mathcal{D}^{\rm s}$ were used and one model was trained for each of
them. Because only numerical gradients were available at the UCCSD(T)
level of theory, weighting force contributions in the loss was reduced
to $\omega_F = 1$. Hence, the focus was put more on validations of
energetics and observables that can be derived from
them. Nevertheless, test set errors on forces and harmonic frequencies
(see Table~\ref{sitab:harmfreq_heh2p}) were still determined.\\

\begin{figure}[ht]
\centering
\includegraphics[width=1.0\textwidth]{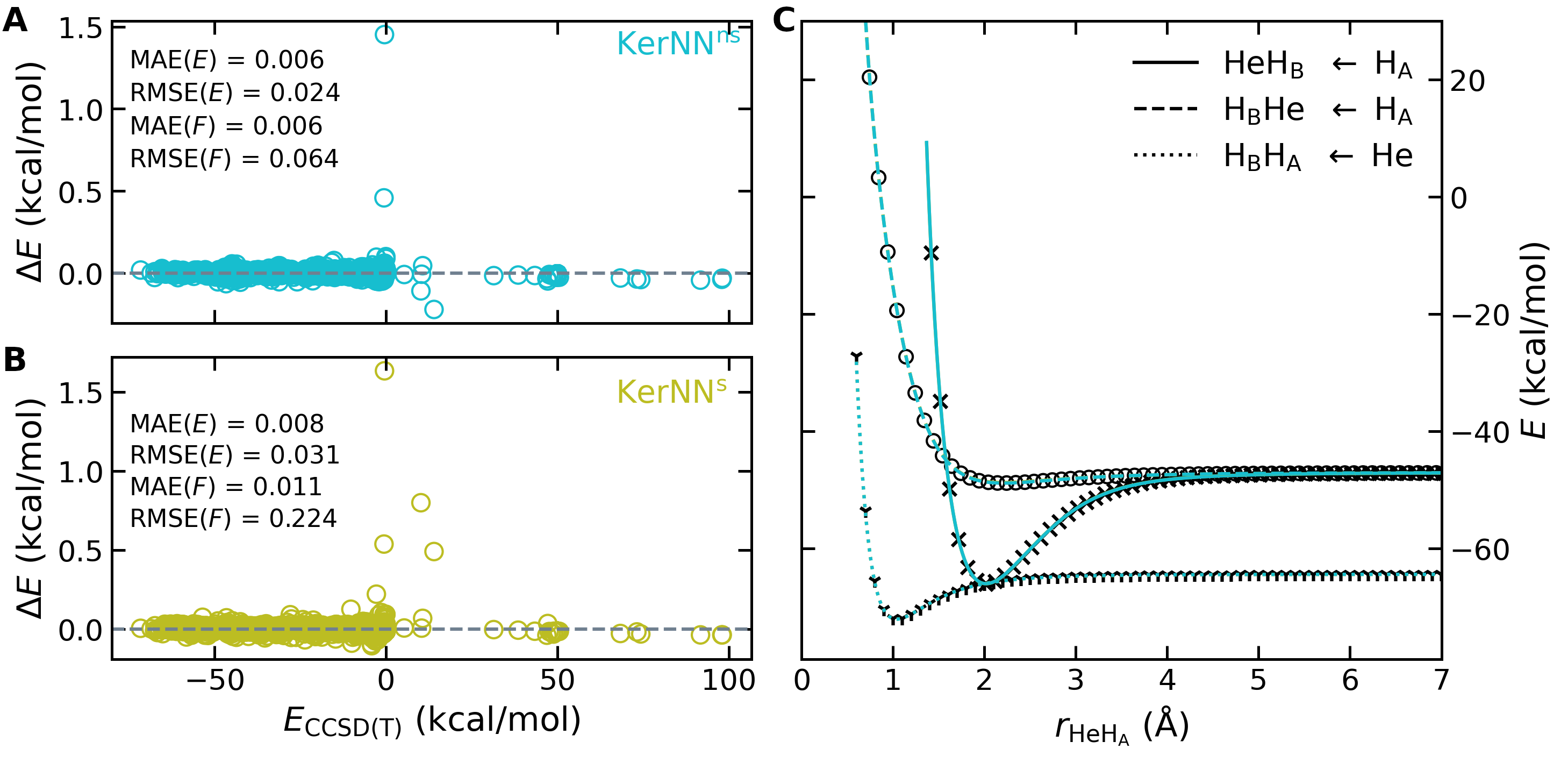}
\caption{Out-of-sample errors for the HeH$_2^+$ KerNN$^{\rm ns}$ (A,
  cyan) and KerNN$^{\rm s}$ (B, olive) PESs (with $\mathcal{D}^{\rm
    ns}$ and $\mathcal{D}^{\rm s}$ as descriptors, respectively)
  trained on UCCSD(T)/aug-cc-pV5Z level reference data. The test set
  contains 4709 randomly chosen structures. Most energies are
  predicted with errors well below 0.1~kcal/mol, while single outliers
  exist (see text). C: One-dimensional potential energy scan along
  $r_{\rm HeH_A}$ for a collinear approach of the reactants as
  determined on KerNN$^{\rm ns}$ (cyan). HeH$_{\rm B}$ $\leftarrow$
  H$_{\rm A}$ refers to diatomic HeH$_{\rm B}$ at its optimized bond
  length with H$_{\rm A}$ approaching atom H$_{\rm B}$. The scans for
  KerNN$^{\rm s}$ are shown, too, but are overlapped by the
  KerNN$^{\rm ns}$ results. Three different atomic arrangements are
  shown and \textit{ab initio} energies determined at the
  UCCSD(T)/aug-cc-pV5Z are illustrated as black symbols. The zero of
  energy was taken with respect to the free atoms.}
\label{fig:kernn_oos_errors_heh2+}
\end{figure}

\noindent
Following the standard practice for assessing the performance of
ML-based PESs, the test set errors were evaluated. From the full
reference data set, 4709 random structures were excluded from the
training. The energy and force prediction errors together with their
averages on the test set for KerNN$^{\rm ns}$ and KerNN$^{\rm s}$ are
shown in Figures~\ref{fig:kernn_oos_errors_heh2+}A and B. The energies
are predicted with high fidelity (${\rm MAE}(E) < 0.01$~kcal/mol for
both KerNN variants). KerNN$^{\rm s}$ has slightly larger averaged
errors than KerNN$^{\rm ns}$, which is likely due to the variability
in the training process. While most predictions were within
0.1~kcal/mol of their \textit{ab initio} reference values, single
outliers with $0.5 < \Delta E < 2.0$ kcal/mol exist. The three
structures with the largest errors are shown in
Figure~\ref{sifig:outliers_heh2p}. Two out of three outliers are the
same in KerNN$^{\rm ns}$ and KerNN$^{\rm s}$ and both structures
feature large interatomic distances ($r(\rm{He-H}) \geq 2.3$~\AA) with
comparable H-He-H angle. Structures close to full dissociation are
likely to pose a challenge for single reference methods such as
UCCSD(T). This can be verified by comparing energies from UCCSD(T) and
FCI calculations for selected structures. It is noted that for a
three-electron system CCSDT and FCI are equivalent which implies that
UCCSD(T) is close but not identical to these highest-accuracy
methods. The two structures considered were the equilibrium
configuration determined on KerNN$^{\rm ns}$ (linear [He--H-H]$^+$)
and the structure with the largest $\Delta E$ between reference
calculations and KerNN$^{\rm ns}$ which is partially dissociated. At
the minimum the energy difference $\Delta_{\rm UCCSD(T) / FCI} = 0.02$
kcal/mol compares with $\Delta_{\rm UCCSD(T) / FCI} = 1.40$ kcal/mol
for the outlier structure. This is a clear indication for the
difficulty of UCCSD(T) to describe geometries close to dissociation
correctly. It is interesting to note that the prediction error of
KerNN is $\sim 1.5$~kcal/mol (see
Figure~\ref{fig:kernn_oos_errors_heh2+}A and B), which is comparable
to $\Delta_{\rm UCCSD(T) / FCI}$. This also implies that KerNN can
identify irregularities in the data set.\\

\noindent
Figure~\ref{fig:kernn_oos_errors_heh2+} (C) shows one-dimensional
potential energy scans along $r_{\rm HeH_{\rm A}}$ as determined on
the KerNN PESs for three different collinear approaches. The
one-dimensional potential energy scans are shown for the KerNN PESs
using $\mathcal{D}^{\rm ns}$ and $\mathcal{D}^{\rm s}$. The results
(cyan and olive lines) are so close that they overlap on the scale of
the plot. Black symbols represent explicit \textit{ab initio} UCCSD(T)
energies using the same grid as for scanning the KerNN PESs. The
exceptional quality of the KerNN PESs is apparent both, for the short
and the long-range part of the PES. The maximum differences between
the PESs and the \textit{ab initio} reference energies for all scans
are $< 0.1$~kcal/mol and emphasize the accuracy of the PESs. In
addition to probing the bond formation process in one dimension in
Figure~\ref{fig:kernn_oos_errors_heh2+}, the accuracy of the reactive
KerNN$^{\rm ns}$ PES, in particular for the He + H$_2^+$ and HeH$^+$ +
H channels, is shown in Figures~\ref{sifig:kernn_2dpes_heh2+} and
\ref{sifig:kernn_2dpes_heh2+_diff}. The comparison to direct UCCSD(T)
calculations yield MAE($E$) of 0.05 and 0.70~kcal/mol for the He +
H$_2^+$ and HeH$^+$ + H channels, respectively. Excluding the ten
predictions with the largest error for the latter (located exclusively
in the repulsive region) yields a MAE($E$) of 0.06~kcal/mol. KerNN is
also suitable to investigate chemical reactions, see
Figure~\ref{sifig:reac_traj_heh2+}. Here two exemplary trajectories,
one non-reactive (A) and the other one reactive (B) using KerNN$^{\rm
  ns}$ are shown. \\

\noindent
As an application of the KerNN PESs the ro-vibrational energies for
the He--H$_2^+$ ionic complex were determined for the KerNN$^{\rm ns}$
PES using the DVR3D suite of programs\cite{tennyson2004dvr3d} which
solves the three-dimensional Schr\"odinger equation in a discrete
variable representation (DVR) for a given PES. Due to the considerable
stabilization of the He--H$_2^+$ ionic complex it is well approximated
as a semi-rigid system with vibrational modes ($\nu_1, \nu_2, \nu_3
\equiv\nu_r, \nu_b, \nu_s$). In a local mode picture, these are the
H-H$^+$ stretch, He--H-H$^+$ bend and the He--H$_2^+$ stretch modes,
respectively.\\

\noindent
Ro-vibrational bound states were determined for 
\textit{ortho-/para}-HeH$_2^+$ for total angular momentum $J=0,1$ in
$e/f$ parity for the complex. Because $o-$/$p-$H$_2^+$ only populates
odd/even $j-$states, the dissociation limits for the two species
differ by $2B = 59.9$ cm$^{-1}$ at the UCCSD(T) level of theory, where
$B$ is the rotational constant of H$_2^+$. A comprehensive list of all
calculated states considered is given in
Table~\ref{sitab:heh2+_boundstates}, which compares results from using
the KerNN$^{\rm ns}$ PES with earlier calculations performed on a
FCI/aug-cc-pV5Z RKHS PES.\cite{MM.heh2:2019} The difference between the earlier and the
present predictions of the bound state energies are given in
Figure~\ref{sifig:kernn_vs_rkhs_boundstates}. The ground state energies
of \textit{o}- and \textit{p}-H$_2^+$($\nu = 0$, $j=0$)--He determined
on the KerNN$^{\rm ns}$ PES are --1793.3578 and --1793.3585~cm$^{-1}$,
which are within less than 0.5 cm$^{-1}$ from the energies determined
on the FCI PES which are --1793.7632 and
--1793.7639~cm$^{-1}$. Comparing all 64 states from the KerNN$^{\rm
  ns}$ and the RKHS PES it is evident that KerNN$^{\rm ns}$ typically
underestimates the bound state energy in comparison to the FCI PES
(\textit{i.e.}, predicts it at lower wavenumber). While this is not
the case for the states with the lowest $n$ it is invariably the case
for states with $n>4$. Notably, the underestimation increases for
near-dissociative bound states (for increasing $n$). Among all states,
the maximum deviation between KerNN$^{\rm ns}$ and RKHS is $\sim
6$~cm$^{-1}$ with a MAE of 2.5~cm$^{-1}$. \\

\noindent
Several transition frequencies were recently characterized
experimentally for the first time from low-resolution spectra in an
ion trap at 4 K using a free electron laser, see Table
\ref{tab:heh2+_band_origins}.\cite{asvany:2021} With the help of
ro-vibrational calculations on the FCI-RKHS PES\cite{MM.heh2:2019} the
experimentally detected peak at 1840 cm$^{-1}$ was assigned to the
H$_2^+$ stretch whereas the peaks at 695 and 840 cm$^{-1}$ correspond
to the He--H$_2^+$ bend and van der Waals stretch modes. These compare
with computed transition frequencies of 1809/1829, 641, and 729
cm$^{-1}$ from DVR3D calculations using the KerNN$^{\rm ns}$ PES which
differ by only a few cm$^{-1}$ from the bound state calculations using
the RKHS-FCI PES. For the former transition frequency (1809/1829),
there are two states because they couple with the bend and stretch
modes of the complex. When comparing computed and experimentally
reported transition frequencies one must note that the widths of the
measured spectra can be considerable, e.g. $\sim 100$ cm$^{-1}$ for
the H$_2^+$ stretch fundamental, with appreciable uncertainty of $\sim
10$ cm$^{-1}$ in the position of the maximum. Also, there is a
shoulder $\sim 20$~cm$^{-1}$ to the red of the measured fundamental
transition at 1840~cm$^{-1}$, which is nicely captured by the
calculations (\textit{i.e.}, 1829 and 1809). Furthermore, two
low-intensity peaks at 1159 and 1234~cm$^{-1}$ can be identified as
the (0200) and (0020) overtones using the calculations, which are
within 30 and 20~cm$^{-1}$ of the computations, respectively. \\

\begin{table}[h]
\begin{tabular}{llllrrrr}\toprule
  $\nu_1$ & $\nu_2$ & $\nu_3$ &$k'$ & Exp.\cite{asvany:2021} & CCVM/RKHS(FCI)\cite{asvany:2021}& DVR3D/KerNN$^{\rm ns}$(UCCSD(T))\\
  \midrule
0	&	1	&	0	&	1	&	695	&	640 [640]	&	640.7 [640.7]	\\
0	&	0	&	1	&	0	&	840	&	731.6 [731.6]	&	729.1 [729.1]	\\
0	&	2	&	0	&	0	&	1159	&	1136.1 [1134.1]	&	1134.8 [1132.8]	\\
0	&	0	&	2	&	0	&	1234	&	1256.4 [1255.6]	&	1252.1 [1251.3]	\\
1	&	0	&	0	&	0	&	1840	&	1812/1833	&	1808.6/1829.3	\\
\end{tabular}
\caption{Measured\cite{asvany:2021} and calculated\cite{asvany:2021}
  vibrational transition frequencies for HeH$_2^+$. The fundamentals
  are (010), (001), and (100) for the bend, van der Waals stretch, and
  H$_2^+$ stretch modes. The CCVM calculations\cite{asvany:2021} were
  carried out on the RKHS representation of the FCI
  energies\cite{MM.heh2:2019} and the DVR3D calculations used the
  KerNN$^{\rm ns}$ representation of the UCCSD(T) data. Transition
  frequencies are given for ortho-[para-]H$_2^+$ separately. It is
  demonstrated that the differences between ortho- and para-energies
  for the two calculations are identical throughout.}
\label{tab:heh2+_band_origins}
\end{table}

\noindent
Comparing the theoretical predictions carried out on two different
PESs based on different levels of theory (FCI vs. UCCSD(T)), two
different representations (RKHS and KerNN), and using two different
methods for computing bound states (CCVM and DVR3D) shows excellent
agreement. This indicates that both calculations are consistent and
precise. This is very encouraging and lends considerable credibility
to a computational approach for predicting spectroscopic properties of
experimentally challenging systems. The RKHS-FCI PES has also been
successfully used together with wavepacket simulations to characterize
Feshbach resonances in
He--H$_2^+$.\cite{MM.heh2:2023,MM.heh2:2024,MM.heh22:2024} While the
experiment and the theoretical calculations are in reasonable
agreement, this calls for additional high-resolution experiments, for
which the theoretical predictions can serve as valuable benchmark.\\

\subsection*{Hydrogen Oxalate}
\begin{figure}[ht]
\centering
\includegraphics[width=0.75\textwidth]{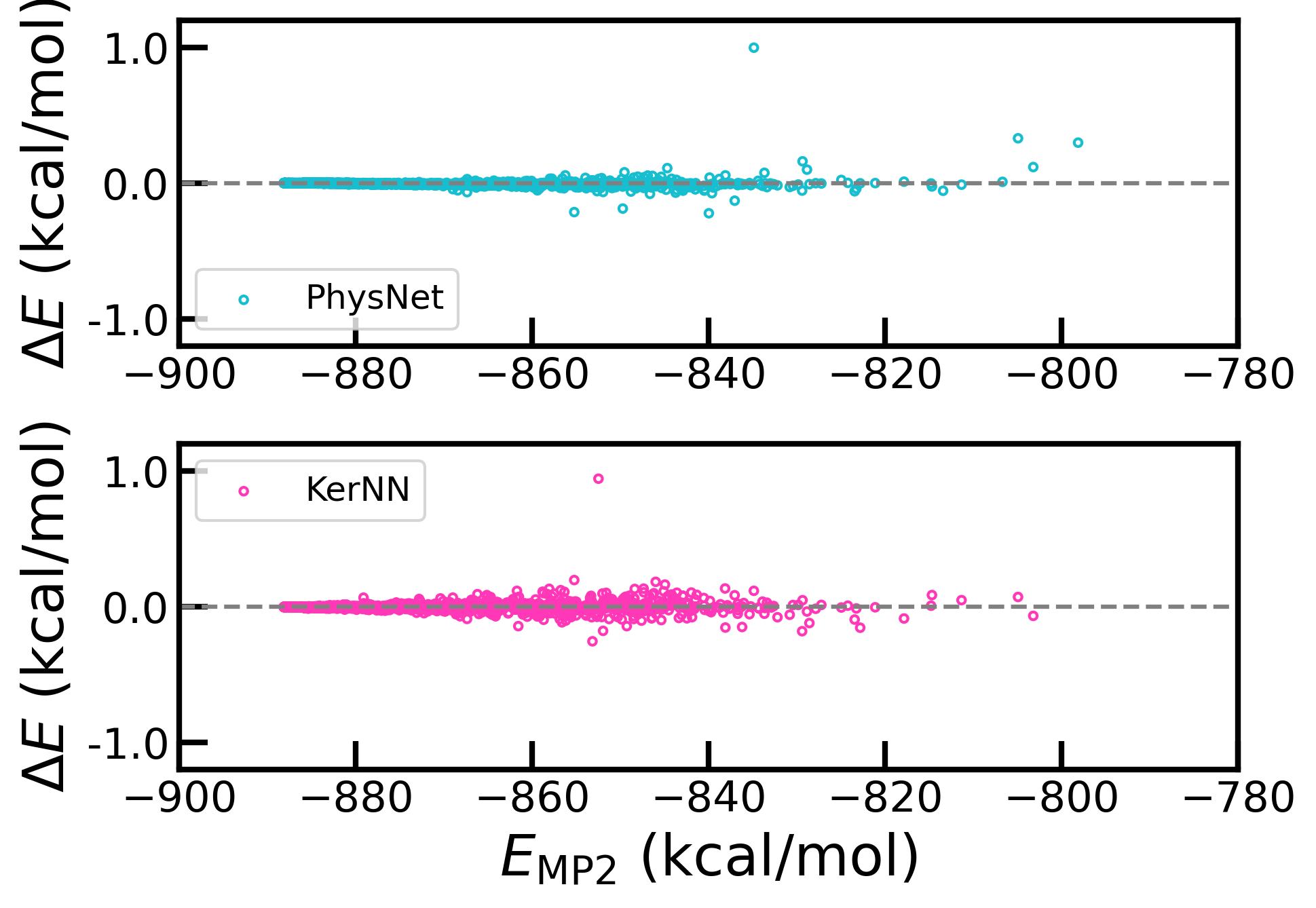}
\caption{Out of sample errors on a test set containing $2000$
  hydrogen oxalate structures for the KerNN$^{\rm ns}$ and PhysNet PESs.}
\label{fig:test_set_errors_hoxa}
\end{figure}

As a final application of KerNN the spectroscopy and reaction dynamics
of hydrogen oxalate are followed. The infrared spectroscopy of hydrogen oxalate has
been characterized from both, experiments and
computations.\cite{wolke2015diffuse,MM.oxa:2017} The system is highly
symmetric, has a strong intramolecular hydrogen bond and features a
corresponding hydrogen transfer. The system serves as a test case to
assess the feasibility for larger, reactive systems. Therefore, only a
KerNN$^{\rm ns}$ was trained and a corresponding PhysNet model was
generated based on the same training data. The quality of the two PESs
on the corresponding test sets is reported in
Figure~\ref{fig:test_set_errors_hoxa} and MAE and RMSE are summarized
in Table~\ref{sitab:oxa1}. It is found that across the full range of
energies covered the KerNN$^{\rm ns}$ PES is competitive with PhysNet
despite the fact that the number of free parameters for PhysNet is
larger by more than two orders of magnitude. In fact, for certain
properties (RMSE$(E)$, RMSE$(F)$, MAE$(\mu)$, and RMSE$(\mu)$)
KerNN$^{\rm ns}$ performs better than PhysNet.\\

\noindent
To study the spectroscopy and dynamics of hydrogen oxalate, MD simulations
were carried out using ASE\cite{larsen2017atomic} for both KerNN$^{\rm
  ns}$ and PhysNet. A total of 100 simulations were started from the
same initial conditions for each of the energy functions. Initially,
the molecule was optimized and random momenta were drawn from a
Maxwell-Boltzmann distribution corresponding to 300~K and assigned to
each of the atoms. The MD simulations were propagated in the $NVE$
ensemble using the velocity Verlet algorithm with a time step of
0.2~fs for 10$^6$ steps. This aggregates to a total of 20~ns
simulation time for each of the PESs.\\

\begin{figure}[h]
\centering \includegraphics[width=1.0\textwidth]{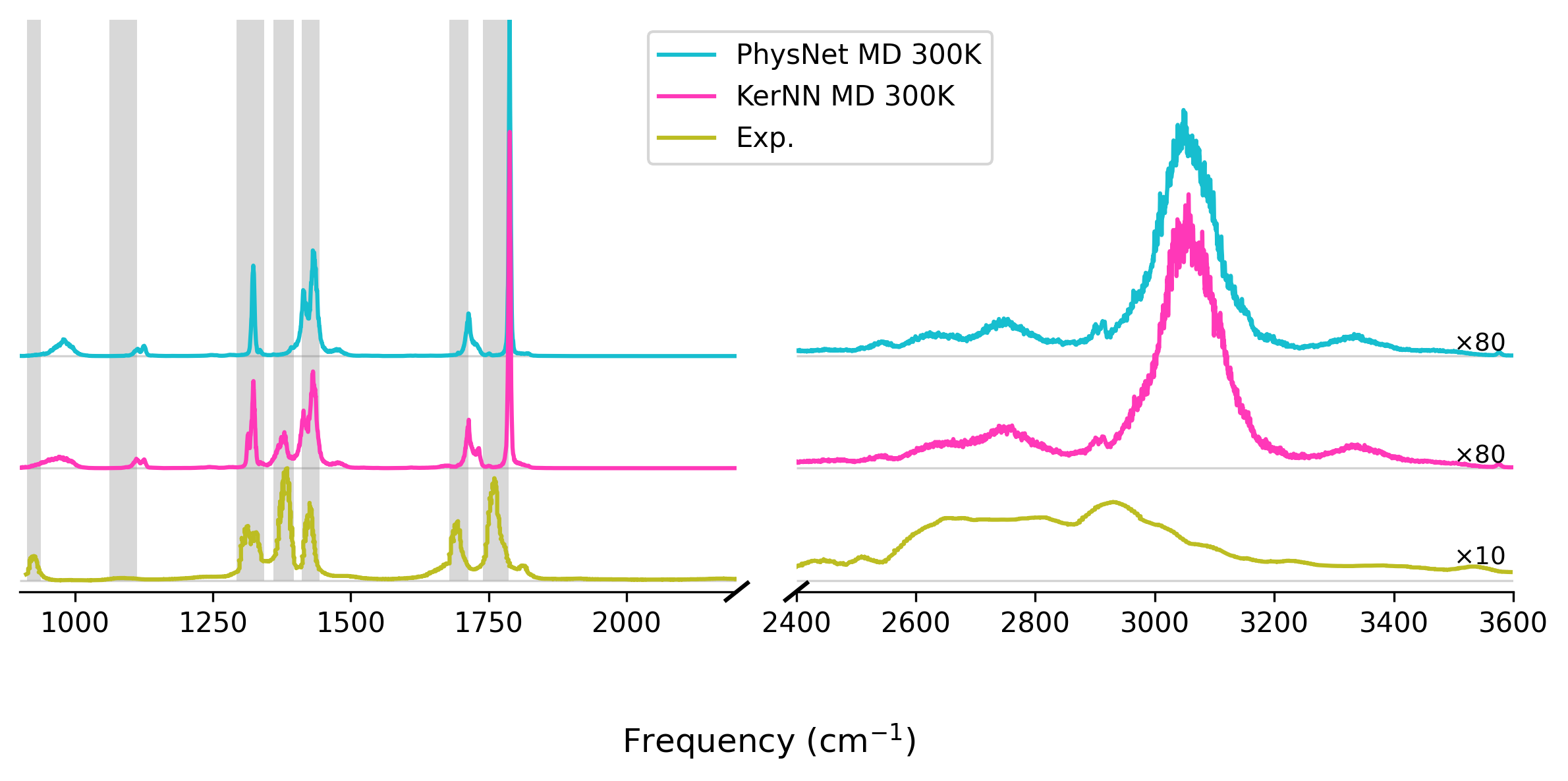}
\caption{Infrared spectra of hydrogen oxalate. The computed spectra
  (top two traces) are obtained from the Fourier transform of the
  dipole-dipole autocorrelation function for MD simulations run with
  KerNN$^{\rm ns}$ and PhysNet. The bottom trace corresponds to an
  experimentally determined (H$_2$-tagged) gas phase
  spectrum\cite{wolke2015diffuse}. The grey-shaded areas correspond to
  the experimentally determined peak positions. }
\label{fig:ir_hoxa_physnet}
\end{figure}

\noindent
The infrared spectra determined from the MD simulations described
above are reported in Figure \ref{fig:ir_hoxa_physnet}. The framework
modes below 2000 cm$^{-1}$ were all captured rather accurately
compared with experiment. It should be noted that the PESs are based
on the MP2 level of theory and that the experiments were carried out
using the H$_2$-messenger technique which, strictly speaking, is not a gas
phase spectrum.  Nevertheless, the interaction between the
H$_2$-messenger and hydrogen oxalate is still weak so that only small
perturbations are expected. A notable feature is that the two computed
spectra from using PhysNet and KerNN (cyan and magenta traces, respectively)
are rather close to one another, except for a peak at
1380~cm$^{-1}$ which appears only for KerNN$^{\rm ns}$ but is
consistent with experiments (olive trace). In other words, the
KerNN$^{\rm ns}$ PES is demonstrably of the same quality as the
PhysNet model - if not even superior. The relative intensities between
experiment and simulations are probably influenced by the fact that
experimentally, a H$_2$ tagging technique was employed whereas the
simulations are in the gas phase. For the most interesting spectral
feature below 3000 cm$^{-1}$ the two simulations agree as well and
allow to assign the measured/calculated pronounced maximum around
2930/3050 cm$^{-1}$ to the O--H stretch or proton transfer mode. In
the experiment, a characteristic plateau with signals between 2600 and
2900~cm$^{-1}$ was detected. This signal range is also detected in the
simulations, albeit with a different intensity pattern. This is in
contrast to earlier work on the vibrational spectroscopy and dynamics
of hydrogen oxalate, which was based on a force field and was unable to
reveal the width of the plateau.\cite{MM.oxa:2017} This is an
advantage of ML-based PESs over traditional force fields as the former
include all inter-mode couplings by learning them directly from
\textit{ab initio} reference data.\\

\noindent
Finally, the same MD simulations that were employed to determine the
IR spectra were used to calculate the hydrogen transfer rates. The barrier height for
hydrogen transfer from reference MP2/aug-cc-pVTZ calculations is 2.355 kcal/mol which
is closely reproduced by KerNN$^{\rm ns}$ (2.360 kcal/mol) and PhysNet
(2.355 kcal/mol), respectively. From an aggregate of 20 ns MD
simulation (100 independent simulations each 200 ps in length), 427
and 524 hydrogen transfers were observed from simulations using the
KerNN$^{\rm ns}$ and PhysNet PESs. This corresponds to transfer rates
of 21/ns and 26/ns. Malonaldehyde, which also features intramolecular
hydrogen transfer between two adjacent O-atoms and is very similar to
hydrogen oxalate, features a H-transfer barrier of 2.79~kcal/mol at the
MP2/aug-cc-pVTZ level. The proton transfer dynamics of malonaldehyde
has been studied in Reference~\citenum{mm.ht:2020}, which suggested
that an O-O motion is gating the H-transfer. At 300~K, a transfer rate
of 7.6/ns was reported. Acetoacetaldehyde/acetylacetone (both being
methyl-substituted variants of malonaldehyde) which was also studied
in earlier work has a H-transfer barrier of 2.59/2.17~kcal/mol at the
MP2/aug-cc-pVTZ level, for which a transfer rate of 15/36 per ns was
found and is consistent with the present findings.\\

\noindent
As long as the hydrogen oxalate remains in its hydrogen-bonded form and the
hydrogen atom transfers between the two oxygen atoms O$_{\rm A}$ and
O$_{\rm B}$ (see Figure \ref{fig:schematics_molecules}) for which
training data is available the simulations are robust and
meaningful. However, since the hydrogen-bond can break at elevated
temperatures and a rotation about the C$_{\rm A}$--C$_{\rm B}$ bond is
possible, exchange of the hydrogen atom to O$_{\rm C}$ or O$_{\rm D}$
also becomes possible. Since the features are not symmetrized, the
trajectory breaks down in such situations. This was not the case for
MD simulations at 300~K. One remedy would be to cover the relevant
symmetries with additional data ("data augmentation") or by including
full or approximate permutational invariance into the model.\\

\section*{Conclusion and Outlook}
The present work introduces KerNN to represent reactive and
non-reactive potential energy surfaces. KerNN capitalizes on the idea
that one-dimensional reproducing kernels provide accurate and
asymptotically meaningful representations suitable for featurization
of a (small) neural network. The main reason to explore such a
combination was the realization that typical present-day NN-based
PESs are over-parametrized by several orders of magnitude. It was
shown in the present case that a KerNN with $\sim 10^3$ performs on
par with a PhysNet model with $\sim 10^6$ parameters trained on the
same reference data set. Occam's razor and the principle of parsimony
state that among competing explanations the simplest explanation with
the fewest variables should be
favoured.\cite{blumer1987occam,domingos1999role} Such considerations also lead other fields, e.g. natural language processing\cite{marah2024miniphi} or computer vision,\cite{lequoc2019efficientnet} to consider the performance of smaller NN-architectures as has been done for KerNN in the present work.
Apart from
being resource-efficient in terms of training times, evaluation times,
memory requirements and reduced concomitant energy consumption, small
NN-based models may also offer advantages when moving towards
explainable and interpretable models (XAI). \\

\noindent
Using symmetrized and unsymmetrized descriptors, $\mathcal{D}^{\rm s}$
and $\mathcal{D}^{\rm ns}$, offers advantages and disadvantages. For
example, for symmetric structures of a symmetric molecule (H$_2$CO)
forces on symmetry-equivalent atoms are exactly the same if the
descriptor is symmetric. On the other hand the training and inference
of KerNN$^{\rm s}$ is computationally more demanding than for
KerNN$^{\rm s}$ and the differences will be more pronounced in larger
systems. This and the (almost) identical performance of the two models
for H$_2$CO raises the question if a rigorous inclusion of physical
constraints (such as the exact permutational invariance of like atoms)
is mandatory. Recent work on the effect on broken symmetries, in
particular for rotations, in ML come to similar conclusions and find
that ``[..]Despite the unquestionable appeal of incorporating
fundamental physical concepts in the architecture of machine-learning
models, it might be beneficial – and it certainly is not as
detrimental one would expect – to just let models
learn[..]''\cite{ceriotti2024brokensym} For KerNN applied to larger
molecules, it will therefore be interesting to assess whether
approximate permutational invariance, \textit{e.g.} by data
augmentation or by including permutational invariance only for
``physically accessible'' permutations, offers advantages over a more
rigorous inclusion of permutational invariance. In addition, it may be
possible to use kernels for describing the atomic environment to
further generalize KerNN.\\

\noindent
The extrapolation capabilities of KerNN were demonstrably
excellent. This is a particular advantage for investigating
bond-breaking and bond-formation. This property is also very
advantageous for reference data from methods that are known to feature
unpredictable convergence, in particular in the asymptotic regions of
the PES, such as multi-reference CI
\cite{MM.heh2:2019,MM.n2o:2020,kon18:094305} or if experimental data
is available.\\

\noindent
A particular advantage over Python-based methods is the fact that
KerNN can be implemented in Fortran which (usually) is computationally
more efficient. This is particularly relevant for long-time
simulations of large systems, which require a large number of
sequential force predictions. As all other NN-based methods, no
explicit analytical form of the underlying model is assumed and
coupling between internal degrees of freedom are explicitly
incorporated into the model. This is of particular interest when
energy transfer phenomena are studied. Future extensions of KerNN
include the combination with accurate electrostatic models such as
MDCM, fMDCM or kMDCM or multipolar electrostatics for modeling
condensed phase systems.\\

\noindent
In summary, the present work introduces an efficient and accurate
approach to represent reactive and non-reactive molecular PESs. KerNN
was applied to infrared spectroscopy and H-transfer reactions for
systems with up to 6 heavy atoms. Extensions to larger molecules are
possible but may require modifications in the kernels used. The most
important insight of the present work is that considerably smaller and
simpler ML-based models can be conceived without compromising their
accuracy.\\

%TC:ignore
\section*{Methods}
This section describes the reference data generation, followed by the
NN architecture and the descriptors used.\\

\subsection*{\textit{Ab initio} Data} 
\textbf{H$_2$CO:} The \textit{ab initio} reference data for H$_2$CO
was available from previous work\cite{MM.h2co:2020}. The data set
contains a total of 4001 configurations generated using normal mode
sampling\cite{smith2017ani} including the optimized H$_2$CO
structure. \textit{Ab initio} energies, forces and dipole moments were
obtained for all structures at the CCSD(T)-F12B/aug-cc-pVTZ-F12 level
of theory using MOLPRO\cite{MOLPRO}. To capture the equilibrium, room
temperature, and higher energy regions of the PES, the normal mode
sampling was carried out at eight different temperatures (10, 50, 100,
300, 500, 1000, 1500, and 2000 K). For each temperature, 500
structures were generated. For training, the energy is taken with
respect to free atoms.\\

\noindent
\textbf{HeH$_2^+$:} The reference structures for the HeH$_2^+$ system
were generated on a grid using Jacobi coordinates ($r$, $R$, $\theta$)
where $r$ is the bond length between diatomic species (either H$_2^+$
or HeH$^+$), $R$ is the distance between the center of mass of the
diatom and the third atom (either He or H) and $\theta$ is the angle
between $\Vec{r}$ and $\Vec{R}$. The details for the radial and
angular grids for the reference structure generation are given in
Table~\ref{sitab:heh2plus_grid}. Structures with any interatomic
distance smaller than 0.6~\AA\/ were discarded. These are complemented
with 500 structures obtained from running $NVT$ MD simulations at
1500~K using the semiempirical GFN2-xTB method\cite{bannwarth2019gfn2}
for each of the diatomic species (the third atom was placed
sufficiently far away). \textit{Ab initio} energies and forces were
determined at the UCCSD(T)/aug-cc-pV5Z level of theory for all
structures. Since only numerical gradients were available for the
UCCSD(T) method in MOLPRO, less relative weight is given to the forces
during training (see Table~\ref{sitab:nn_hyperparams}). Structures with
energies 100~kcal/mol or higher than the complete dissociation
(\textit{i.e.}, [He + H + H]$^+$) were excluded, yielding a total of
62834 structures. The zero of energy was taken with respect to the
free atoms (\textit{i.e.}, [He + H + H]$^+$ is at 0 kcal/mol).\\

\noindent
\textbf{Hydrogen oxalate:} Structures for hydrogen oxalate were
sampled by running $NVT$ MD simulations at multiple temperatures (100,
300, 500, 1000 and 1500~K) using the semiempirical GFN2-xTB
method\cite{bannwarth2019gfn2} (2500 structures each except for 1500~K
for which only 1000 geometries were generated) as implemented in 
ASE.\cite{larsen2017atomic} The region
around the proton transfer transition state was sampled with an
artificial harmonic potential (1500 structures) and at $T =
500$~K. Additionally, normal mode sampling\cite{smith2017ani} at
increasing temperatures (100, 300, 500, 100, 1500, 2000~K) was carried
out for both the optimized and transition state structure of hydrogen
oxalate (800 per $T$). The data set was augmented based on adaptive
sampling and diffusion Monte Carlo
simulations\cite{kosztin1996introduction,kaser2022transfer} using
PhysNet\cite{MM.physnet:2019} to ensure robustness of the PES. The
final data set contained a total of 22110 structures, for which
energies, forces and dipole moments were determined at the
MP2/aug-cc-pVTZ level of theory using MOLPRO\cite{MOLPRO}. Again, the
energy was taken with respect to free atoms. \\

\subsection*{Neural Network}
The PESs for the three molecular systems were represented by a small
and fully connected feed-forward NN. The fundamental building blocks
of NNs are dense layers of the form
\begin{align}
    \bm{y} =\sigma (\bm{Wx} + \bm{b}),
\end{align}
which need to be stacked and combined with a non-linear activation
function $\sigma$ to model non-linear relationships within the
data. Here, $\bm{x} \in \mathbb{R}^{n_{\rm in}}$ ($\bm{y}\in
\mathbb{R}^{n_{\rm out}}$) corresponds to the input (output) vector
with a dimensionality of $n_{\rm in}$ ($n_{\rm out}$) and the
activation is applied entry-wise. $\bm{W} \in \mathbb{R}^{n_{\rm out}
  \times n_{\rm in}}$ and $\bm{b}\in \mathbb{R}^{n_{\rm out}}$ are the
weights and biases that are learnable parameters. The NN architecture
used throughout this work was
\begin{align}
\label{eq:ffnet}
    V = \bm{W_3}\sigma(\bm{W_2}\sigma(\bm{W_1}\sigma(\bm{W_0 k(r)} +
    \bm{b_0}) + \bm{b_1}) + \bm{b_2}) + b_3 ,
\end{align}
and consists of an input (0), two hidden (1 and 2), and an output (3)
layer. The activation function $\sigma$ was a
soft plus function, and the last layer was a linear
transformation. The input to the NN are functions of the interatomic
distances $\bm{k(r)}$ (\textit{vide infra}) and the output is the
potential energy $E$ of the molecule. The forces acting on the atoms
were obtained from reverse mode automatic
differentiation.\cite{baydin2017automatic}\\

\noindent
The learnable parameters of the NN were optimized by minimizing an
appropriate loss function $\mathcal{L}$ using
AMSGRAD.\cite{kingma2014adam} Here, a mean squared error loss was used
and the loss function was
\begin{align}\label{eq:loss_kernn}
    \mathcal{L} = \left|E - E^{\rm ref}\right| + \omega_F \sum_{i=1}^N
    \sum_{\alpha=1}^3\left| - \frac{\partial E}{\partial x_{i,\alpha}}
    - F^{\rm ref}_{i,\alpha} \right|.
\end{align}
$E$ and $E^{\rm ref}$ correspond to the model and reference energies,
$F^{\rm ref}_{i,\alpha}$ are the Cartesian components of the reference
forces on atom $i$, and $x_{i,\alpha}$ is the $\alpha$th Cartesian
coordinate of atom $i$. $\omega_F$ is a hyperparameter to weigh the
contribution of the forces to the total loss function. During
training, an exponential moving average of all learnable parameters
was tracked using a decay rate of 0.999. Overfitting was prevented
using early stopping: After every epoch, the loss function was
evaluated on a validation set of reference structures using the
parameters obtained from the exponential moving
average.\cite{prechelt2002early} After training, the model that
performed best on the total validation loss
(Equation~\ref{eq:loss_kernn} or \ref{eq:loss_kernn_dipole} depending
on whether dipole moments are required) was selected.\\

\noindent
If a dipole moment surface was desired, \textit{e.g.}, for
spectroscopic studies, the same NN can be used with only little
adaptation. First, the number of output nodes needs to be changed to
$N_{\rm atoms} + 1$ (one for the energy and one each for atomic
partial charges $q_i$). Inspired by PhysNet\cite{MM.physnet:2019} (and
similar NNs) and given that the dipole moment is $\bm{\mu} =
\sum_{i=1}^{N_{\rm atoms}} q_i\bm{x}_{i,}$, the loss function is
changed to
 \begin{align}\label{eq:loss_kernn_dipole}
    \mathcal{L} = \left|E - E^{\rm ref}\right| + \omega_F \sum_{i=1}^N
    \sum_{\alpha=1}^3\left| - \frac{\partial E}{\partial x_{i,\alpha}}
    - F^{\rm ref}_{i,\alpha} \right| + \omega_\mu \left|
    \sum_{\alpha=1}^3\sum_{i=1}^N q_i x_{i,\alpha} - \mu_\alpha^{\rm
      ref} \right| + \omega_Q \left|\sum_{i=1}^N q_i - Q^{\rm ref}
    \right|
\end{align}
where $q_i$ is the partial atomic charge of atom $i$, $Q^{\rm ref}$ is
the total charge of the system and $\omega_\mu$ and $\omega_Q$ are the
corresponding hyperparameters.\\

\subsection*{Descriptors}
The design of molecular descriptors (\textit{i.e.}, encoding a
molecular configuration in a machine-readable format) is a pertinent
problem in quantum ML and an active area of
research.\cite{raghunathan2022molecular} Such descriptors ideally
satisfy several key criteria including (i) invariance under
transformations such as translation, rotation, and permutation of
identical elements, (ii) uniqueness, exhibiting changes when
transformations that alter the predicted properties are applied, and
(iii) continuity and differentiability with respect to atomic
coordinates to enable the calculation of forces in molecular
simulations.\cite{kaser2023neural} Descriptors can typically be
categorized into two groups: fixed or learnable
descriptors.\cite{kocer2022neural}\\

\noindent
While translational and rotational invariance are straightforward to
satisfy by resorting to internal coordinates, permutational invariance
is more challenging to achieve. Many high-dimensional NNs obtain
atomic contributions to a total energy, which satisfies the
permutational invariance by
construction.\cite{behler2007generalized,MM.physnet:2019} Alternative
routes to permutationally invariant PESs include data
augmentation\cite{zou2003new}, symmetrization of
NNs,\cite{prudente1998fitting,nguyen2012modified} or input
symmetrization.\cite{jiang2013permutation,shao2016communication} The
latter is frequently achieved by using a permutation invariant basis
to generate permutationally invariant polynomials, which were applied
successfully to numerous
systems.\cite{braams2009permutationally,xie2010permutationally,bowman2011high}
However, including the permutational invariance into a PES usually
requires a large function space for fitting and the transformed,
symmetrized basis will be larger than (or equal to) the original
vector space.\\

\noindent
The descriptors used here are based on concepts rooted in the theory
of reproducing
kernels\cite{aronszajn1950theory,ho96:2584,MM.rkhs:2017,MM.rkhs:2020}
which have also been successfully employed for representing reactive
PESs for small
molecules.\cite{kon18:094305,MM.heh2:2019,san20:3927,pez20:2171}
One-dimensional reciprocal power reproducing kernels
\begin{align}\label{eq:1d_kernels}
    k^{[n,m]} (x,x') = n^2x_>^{-(m+1)} B(m+1, n) _2F_1\left(-n + 1,
    m+ 1; n+m+1; \frac{x_<}{x>}\right)
\end{align}
were shown to represent diatomic potentials
faithfully.\cite{ho96:2584,soldan2000long} In Eq.~\ref{eq:1d_kernels}
$x_<$ and $x_>$ correspond to the smaller and larger values of $x$ and
$x'$, respectively, $[n,m]$ are smoothness and asymptotic reciprocal
power parameters, $B(a, b) = \frac{(a-1)!  (b-1)!}{(a+b-1)!}$ is the
$\beta-$function and $_2F_1(a,b;c;z)$ is Gauss' hypergeometric
function.\cite{ho96:2584} The one-dimensional kernels effectively
serve as a similarity function between pairs of values $x$ and $x'$.\\

\noindent
In the present work the $k^{[3,3]}$ kernel
\begin{align}\label{eq:33kernel}
    k^{[3,3]}(r, r') = \frac{3}{20 r_>^4} - \frac{6}{35}
    \frac{r_<}{r_>^5} + \frac{3}{56}\frac{r_<^2}{r_>^6}
\end{align}
was used throughout where $r$ and $r'$ are interatomic distances of a
query structure and a reference structure, respectively. For example,
the reference structure could be an equilibrium configuration,
transition state structure, \textit{etc.}. Again, the values $r_<$ and
$r_>$ are the smaller and larger values of $r$ and $r'$,
respectively. The current approach uses the 1D kernels as local
features to build a global descriptor $\mathcal{D}$ which is then the
input to the NN. Consequently, the method is referred to as KerNN =
``kernels + NN'' in the following.\\

\noindent
{\bf Descriptors for H$_2$CO:} The first, non-symmetrized variant of
KerNN, KerNN$^{\rm ns}$, uses the six one-dimensional kernels directly
as input to the feed-forward NN with the descriptor given by
$\mathcal{D}^{\rm ns}$ in Equation~\ref{eq:descriptor_h2co}. Hence,
the global descriptor $\mathcal{D}^{\rm ns}$ takes translational and
rotational invariance into account but neglects the permutational
invariance of the two equivalent H-atoms.\\

\begin{align}\label{eq:descriptor_h2co}
    \mathcal{D}^{\rm ns} = \begin{bmatrix}
           k^{[3,3]}(r_{\rm CO},r_{\rm CO}')\\
           k^{[3,3]}(r_{\rm CH_{\rm A}},r_{\rm CH_{\rm A}}')\\
           k^{[3,3]}(r_{\rm CH_{\rm B}},r_{\rm CH_{\rm B}}')\\
           k^{[3,3]}(r_{\rm OH_{\rm A}},r_{\rm OH_{\rm A}}')\\
           k^{[3,3]}(r_{\rm OH_{\rm B}},r_{\rm OH_{\rm B}}')\\
           k^{[3,3]}(r_{\rm H_{\rm A} H_{\rm B}},r_{\rm H_{\rm A} H_{\rm B}}')\\
         \end{bmatrix}
         \quad\quad
     \mathcal{D}^{\rm s} = \begin{bmatrix}
           k^{[3,3]}(r_{\rm CO},r_{\rm CO}')\\
           k^{[3,3]}(r_{\rm CH_{\rm A}},r_{\rm CH_{\rm A}}') + k^{[3,3]}(r_{\rm CH_{\rm B}},r_{\rm CH_{\rm B}}')\\
           k^{[3,3]}(r_{\rm OH_{\rm A}},r_{\rm OH_{\rm A}}') + k^{[3,3]}(r_{\rm OH_{\rm B}},r_{\rm OH_{\rm B}}')\\
           k^{[3,3]}(r_{\rm CH_{\rm A}},r_{\rm CH_{\rm A}}')^2 + k^{[3,3]}(r_{\rm CH_{\rm B}},r_{\rm CH_{\rm B}}')^2\\
           k^{[3,3]}(r_{\rm OH_{\rm A}},r_{\rm OH_{\rm A}}')^2 + k^{[3,3]}(r_{\rm OH_{\rm B}},r_{\rm OH_{\rm B}}')^2\\
           k^{[3,3]}(r_{\rm H_{\rm A} H_{\rm B}},r_{\rm H_{\rm A} H_{\rm B}}')\\
         \end{bmatrix}
\end{align}

\noindent
The second variant of KerNN, KerNN$^{\rm s}$, accounts for
permutational invariance of equivalent H-atoms explicitly. The
approach chosen to include permutation invariance into KerNN is
inspired by Reference~\citenum{shao2016communication}, which uses
fundamental invariants\cite{derksen2015computational}. In contrast to
the primary and secondary polynomials in
PIPs\cite{braams2009permutationally} and the polynomials in the PIP-NN
method\cite{jiang2013permutation}, fundamental invariants comprise the
minimum number of invariants necessary to generate all invariant
polynomials. Dedicated software to calculate the fundamental
invariants using King's algorithm exist\cite{king2013minimal,singular}
and exemplary fundamental invariants for a selection of molecules can
be found in Reference~\citenum{shao2016communication}. The
permutationally invariant descriptor $\mathcal{D}^{\rm s}$ used here
is given explicitly in Equation~\ref{eq:descriptor_h2co}. The
interatomic distances of the optimized structure serve as reference
$r_i'$.\\

\noindent
{\bf Descriptors for [HeH$_2$]$^+$:} The PES for HeH$_2^+$ is also
represented by two different global descriptors. The first,
$\mathcal{D}^{\rm ns}$, does not take the permutational invariance
into account, while the second, $\mathcal{D}^{\rm s}$ includes it
using fundamental invariants, see
Equation~\ref{eq:descriptor_heh2p}. The reference structures $r_i'$
for $\mathcal{D}^{\rm ns}$ and $\mathcal{D}^{\rm s}$ differed: for
$\mathcal{D}^{\rm ns}$ it was the linear H$_{\rm A}$-H$_{\rm B}$-He
arrangement with $r'_{\rm H_{\rm A} He} = 2.121$ and $r'_{\rm H_{\rm
    B} He} = 1.021$~\AA\ whereas for $\mathcal{D}^{\rm s}$ a
symmetrized structure is required which was chosen to be the linear
H-He-H arrangement with $r'_{\rm H-He} = 0.9$~\AA\/ were used. One
advantage of using the one-dimensional kernel descriptors for systems
in which the full dissociation of the molecule is conceivable is that
they go to zero for large $r$ (see
Figure~\ref{sifig:kernn_heh2p_1dk}).\\

\begin{align}\label{eq:descriptor_heh2p}
    \mathcal{D}^{\rm ns} = \begin{bmatrix}
           k^{[3,3]}(r_{\rm H_{\rm A} He},r_{\rm H_{\rm A} He}')\\
           k^{[3,3]}(r_{\rm H_{\rm B} He},r_{\rm H_{\rm B} He}')\\
           k^{[3,3]}(r_{\rm H_{\rm A} H_{\rm B}},r_{\rm H_{\rm A} H_{\rm B}}')\\
         \end{bmatrix}
         \quad\quad
     \mathcal{D}^{\rm s} = \begin{bmatrix}
           k^{[3,3]}(r_{\rm H_{\rm A} He},r_{\rm H_{\rm A} He}') + k^{[3,3]}(r_{\rm H_{\rm B} He},r_{\rm H_{\rm B} He}')\\
           k^{[3,3]}(r_{\rm H_{\rm A} He},r_{\rm H_{\rm A} He}')^2 + k^{[3,3]}(r_{\rm H_{\rm B} He},r_{\rm H_{\rm B} He}')^2\\
           k^{[3,3]}(r_{\rm H_{\rm A} H_{\rm B}},r_{\rm H_{\rm A} H_{\rm B}}')\\
         \end{bmatrix}
\end{align}

\noindent
{\bf Descriptors for hydrogen oxalate:} For hydrogen oxalate, only a
non permutationally invariant descriptor was employed. It is
constructed from 21 one-dimensional $k^{[3,3]}$ kernels, one for each
interatomic distance. The optimized hydrogen oxalate structure served as
reference structure.

\section*{Data Availability}
The codes and data for the present study are available from
\url{https://github.com/MMunibas/KerNN} upon publication.

%\bibliography{references}

\section*{Acknowledgment}
The authors gratefully acknowledge financial support from the Swiss
National Science Foundation through grants $200020\_219779$ (MM),
$200021\_215088$ (MM), the NCCR-MUST (MM), the AFOSR (MM), and the
University of Basel (MM).\\

\section*{Competing Interests}
The authors declare no competing interests.

\clearpage
\setcounter{figure}{0}
\setcounter{table}{0}
\renewcommand{\thepage}{S\arabic{page}}
\renewcommand{\thetable}{S\arabic{table}}
\renewcommand{\thefigure}{S\arabic{figure}}
\section*{Supporting Information: The Bigger the Better? Accurate Molecular Potential Energy Surfaces from Minimalist Neural Networks}

\section*{Methods}

\begin{table}[h]
\begin{tabular}{ccccccc}\toprule
&\multicolumn{2}{c}{H$_2$CO}
  &\multicolumn{2}{c}{HeH$_2^+$}&\multicolumn{2}{c}{H. Oxa.}\\\cmidrule(lr){2-3}\cmidrule(lr){4-5}\cmidrule{6-6}
&KerNN$^{\rm ns}$&KerNN$^{\rm s}$&KerNN$^{\rm ns}$&KerNN$^{\rm s}$&KerNN$^{\rm ns}$\\\midrule 
$N^{\rm nodes}_{\rm in}$	&	6	&	7 &	3 & 3	&21	\\
$N^{\rm nodes}_{\rm hidden}$ & 20 & 20 & 40 & 40 & 50	\\
$N^{\rm layers}_{\rm hidden}$ & 2 & 2 & 2 & 2 & 2	\\
$\omega_F$ & 10 & 10 & 1 & 1 & 10	\\
$\omega_\mu$ & 5 & -- & -- & --  & 5	\\
$\omega_Q$ & 2 & -- & -- & -- & 2	\\
$N_{\rm params}$	&	1001 &	1021 & 3481 & 3481 &6608	&	\\
\bottomrule
\end{tabular}
\caption{Parameters and hyperparameters of the neural networks used
  throughout this work.}
\label{sitab:nn_hyperparams}
\end{table}

\begin{table}[H]
\begin{tabular}{ccc}\toprule
Quantity / Diatom     & H$_2^+$ & HeH$^+$\\\midrule
$r$ & 0.5/5/0.2 and 6/50/1 & 0.5/5/0.2\\
$R$ & 0.5/5/0.2 and 6/50/1 & 0.5/5/0.2\\
$\Theta$ & 0/180/15 & 0/180/15\\
$N_{\rm tot}$ & 58357 & 6877\\\bottomrule
\end{tabular}
\caption{Grid for the triatomic HeH$_2^+$ potential. All values are given in \AA\/ and
degrees, respectively. The quantities are given as min/max/step.}
\label{sitab:heh2plus_grid}
\end{table}

\begin{figure}[H]
\centering
\includegraphics[width=0.9\textwidth]{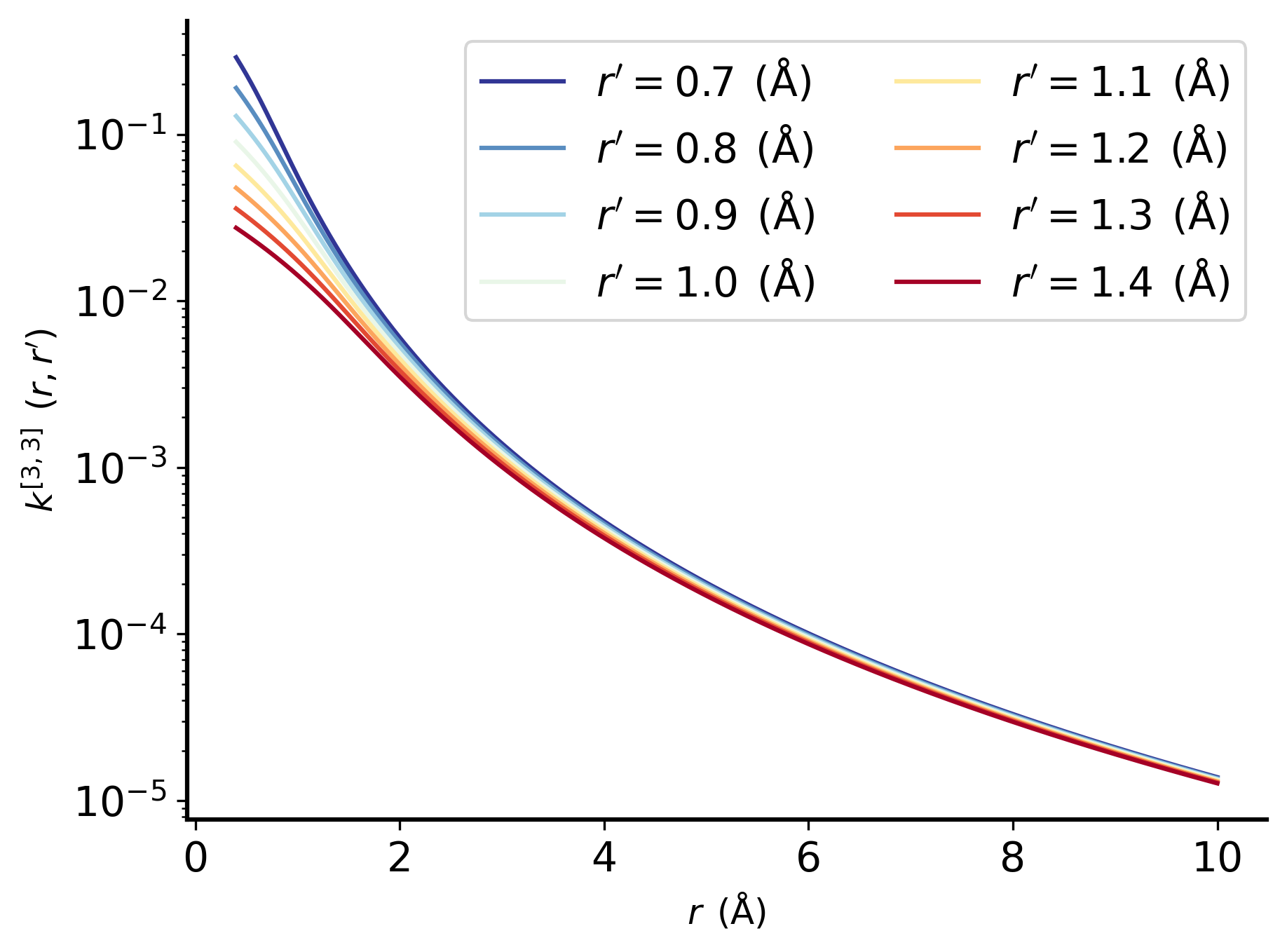}
\caption{The one-dimensional $k^{[3,3]}(r, r')$ kernel as a function of $r$ and for different
reference values $r'$. The curves illustrate the asymptotic behaviour of the curves. Note that the y-axis is displayed on a logarithmic scale.}
\label{sifig:kernn_heh2p_1dk}
\end{figure}

\subsection*{Bound state calculations for HeH$_2^+$}
Ro-vibrational bound states have been calculated for different $J$ (total angular momentum) states with $e$ and $f$ symmetries for both $ortho-$ and $para-$HeH$_2^+$ complex. The DVR3D program\cite{tennyson2004dvr3d} was used to solve the eigenvalue problem. Jacobian coordinates were chosen to represent the spectroscopic system in a 3D discrete variable representation (DVR) formulation. The radial grids along the $R$ and $r$ coordinates were defined by 86 and 32 Gauss-Laguerre quadrature points, respectively and for the angular grid (Jacobi angle $\theta$) 36 Gauss-Legendre points were used. The wavefunctions along $R$ and $r$ were constructed using Morse oscillator functions while the angular part of the wavefunctions was represented by Legendre polynomials. For the $r-$degree of freedom (H$_2^+$), $r_e = 2.5$ a$_0$, $D_e = 0.1026$ $E_{\rm h}$ and $\omega_e$ = 0.018 $E_{\rm h}$ were used whereas for $R$ the values were $R_e = 11.5$ a$_0$, $D_e = 0.08$ $E_{\rm h}$, and $\omega_e = 0.00065$ $E_{\rm h}$. These parameters were chosen so that a large region in the configuration space can be covered by the wavefunction to obtain the near dissociation states. The $r_2-$embedding\cite{tennyson2004dvr3d} was used to compute the $J>0$ states, where the $z$-axis is parallel to $R$ in body-fixed Jacobi coordinates. In the $r_2-$embedding, calculations with $ipar = 1$ and 0 correspond to the $ortho-$ and $para-$H$_2^+$ states, respectively, while the $e$ and $f$ symmetries are defined by the parity operator $p$. Coriolis couplings were included in the Hamiltonian for $J>0$ calculations.\\

\section*{Results and discussion}
\subsection*{H$_2$CO}

\begin{table}[ht]
\centering
\begin{tabular}{rrrrrrr}\toprule
 (cm$^{-1}$) & \textbf{KerNN$^{\rm ns}$} & \textbf{KerNN$^{\rm s}$}& \textbf{RKHS+F} & \textbf{PhysNet}& \textbf{FCHL} & \textbf{CCSD(T)-F12}\\\midrule
1	&	1186.4	&	1186.5	&	1186.4	&	1186.4	&	1186.5	&	1186.5	\\
2	&	1267.7	&	1267.9	&	1268.1	&	1268.2	&	1268.0	&	1268.2	\\
3	&	1532.5	&	1532.6	&	1532.6	&	1532.6	&	1532.7	&	1532.7	\\
4	&	1776.4	&	1776.5	&	1776.4	&	1776.4	&	1776.4	&	1776.4	\\
5	&	2933.4	&	2933.8	&	2933.5	&	2933.6	&	2933.6	&	2933.8	\\
6	&	3006.0	&	3005.6	&	3005.8	&	3005.7	&	3005.6	&	3005.8	\\\midrule
\textbf{MAE}  &  0.2 &   0.1     &  0.1 &  0.1  &  0.1   &     \\\bottomrule  
\end{tabular}\caption{Harmonic frequencies as obtained on the KerNN$^{\rm ns}$ and KerNN$^{\rm s}$ PESs for H$_2$CO in comparison to their CCSD(T)-F12B/aug-cc-pVTZ-F12 reference frequencies. These are compared to three PESs for H$_2$CO based on the RKHS+F\cite{MM.rkhs:2020} method, PhysNet\cite{MM.physnet:2019} and obtained from kernel ridge regression using the FCHL descriptor\cite{faber2018alchemical}, which were reported in Reference~\cite{MM.h2co:2020}. All PESs have been trained on the same reference data set containing a total of 4001 H$_2$CO structures with corresponding energies and forces.}\label{sitab:kernn_harmfreq}
\end{table}

\begin{table}[ht]
\centering
\begin{tabular}{rr}\toprule
& \textbf{KerNN$^{\rm ns}$} \\\midrule
MAE($E$):	&	0.00052	\\
RMSE($E$):	&	0.00090 \\
MAE($F$):	&	0.00902	\\
RMSE($F$):	&	0.01786	\\
MAE($D$):	&	0.00071	\\
RMSE($D$):	&	0.00129	 \\
\end{tabular}\caption{Test set errors for KerNN$^{\rm ns}$ trained on energies, forces and dipole moments using the loss function given in Equation~\ref{eq:loss_kernn_dipole}.  Errors of the energy, forces and dipole moments are given in kcal/mol, kcal/mol/\AA/ and Debye, respectively. }\label{sitab:kernn_h2co_efmu}
\end{table}

\begin{figure}[ht]
\centering
\includegraphics[width=0.8\textwidth]{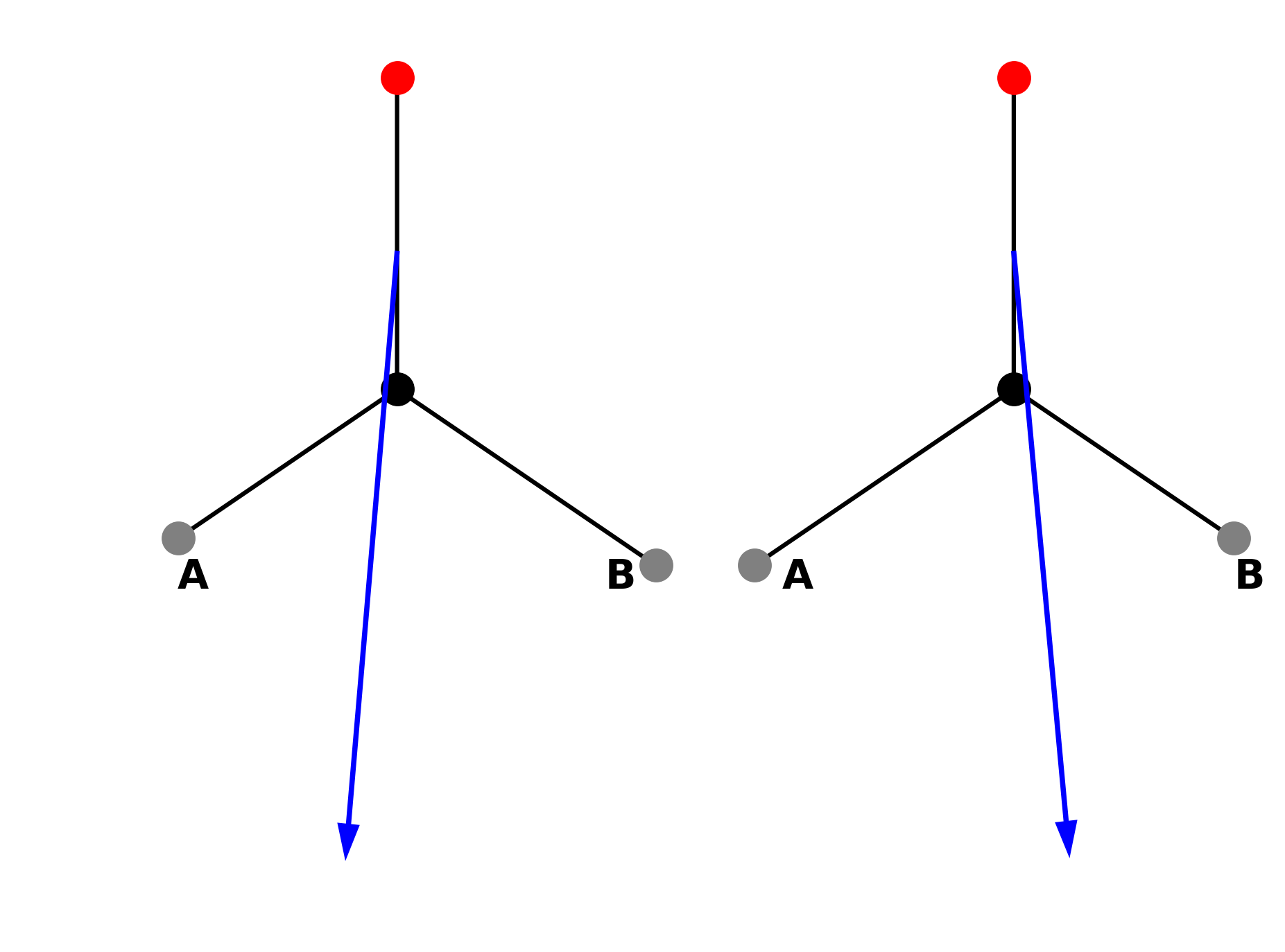}
\caption{Two symmetric H$_2$CO configurations in which two sets of  C-H bond lengths are the same (the C-H$_{\rm A}$ bond length in the left is the same as C-H$_{\rm B}$ in the right configuration). In the permutationally invariant formulation, KerNN$^{\rm s}$, the two configurations are described by the same molecular descriptor $\mathcal{D}^s$ which leads to the same potential energy by construction. Since in the given formulation, the dipole moment (blue) is obtained from $\sum_{\alpha=1}^3\sum_{i=1}^N q_i x_{i,\alpha}$ this would require different partial charges for, \textit{e.g.}, H$_{\rm A}$ (left) and H$_{\rm A}$ (right). As the two structures have the same descriptor and the formulation, therefore, lacks a direct mapping to the partial charges (\textit{i.e.} in this case we would try to predict two sets of partial charges with the same descriptor), it is not possible to learn the dipole moment with KerNN$^{\rm s}$.}
\label{sifig:kernns_h2co_dipole_inability}
\end{figure}

\begin{table}[ht]
\centering
\begin{tabular}{rccccc}\toprule
 (ms)         & \textbf{KerNN$^{\rm ns}$}                                          & \textbf{KerNN$^{\rm s}$}& \textbf{RKHS+F} & \textbf{PhysNet}& \textbf{FCHL} \\\midrule
 time/eval & 1.3/\textbf{0.012}/\textbf{0.003$^*$} & 2.0  & \textbf{0.3}           & 34                           &92\\\bottomrule
\end{tabular}
\caption{Computational timings for KerNN$^{\rm ns}$, KerNN$^{\rm s}$,
  RKHS+F, PhysNet and kernel ridge regression with the FCHL
  descriptor. The reported values correspond to (consecutive) energy
  and force evaluations as would be required in a MD
  simulation. Numbers in plain (boldface) letters correspond to Python
  implementations used via ASE (FORTRAN implementations used via
  CHARMM).$^*$ Pure FORTRAN}\label{sitab:kernn_timings}
\end{table}

\clearpage
\subsection*{HeH$_2^+$}
\begin{table}[]
\begin{tabular}{lll}\toprule
(cm$^{-1}$) & KerNN$^{\rm ns}$  & CCSD(T) \\\midrule
1 & 711.5 & 717.9  \\
2 & 711.5 & 717.9 \\
3 & 989.2 & 986.6\\
4 & 1959.1  & 1935.4\\\bottomrule
\end{tabular}\caption{Harmonic frequencies as obtained on the KerNN$^{\rm ns}$ PES for a linear H-H-He arrangement in comparison to their UCCSD(T)/aug-cc-pV5Z reference frequencies.}\label{sitab:harmfreq_heh2p}
\end{table}

\begin{figure}[h]
\centering
\includegraphics[width=0.8\textwidth]{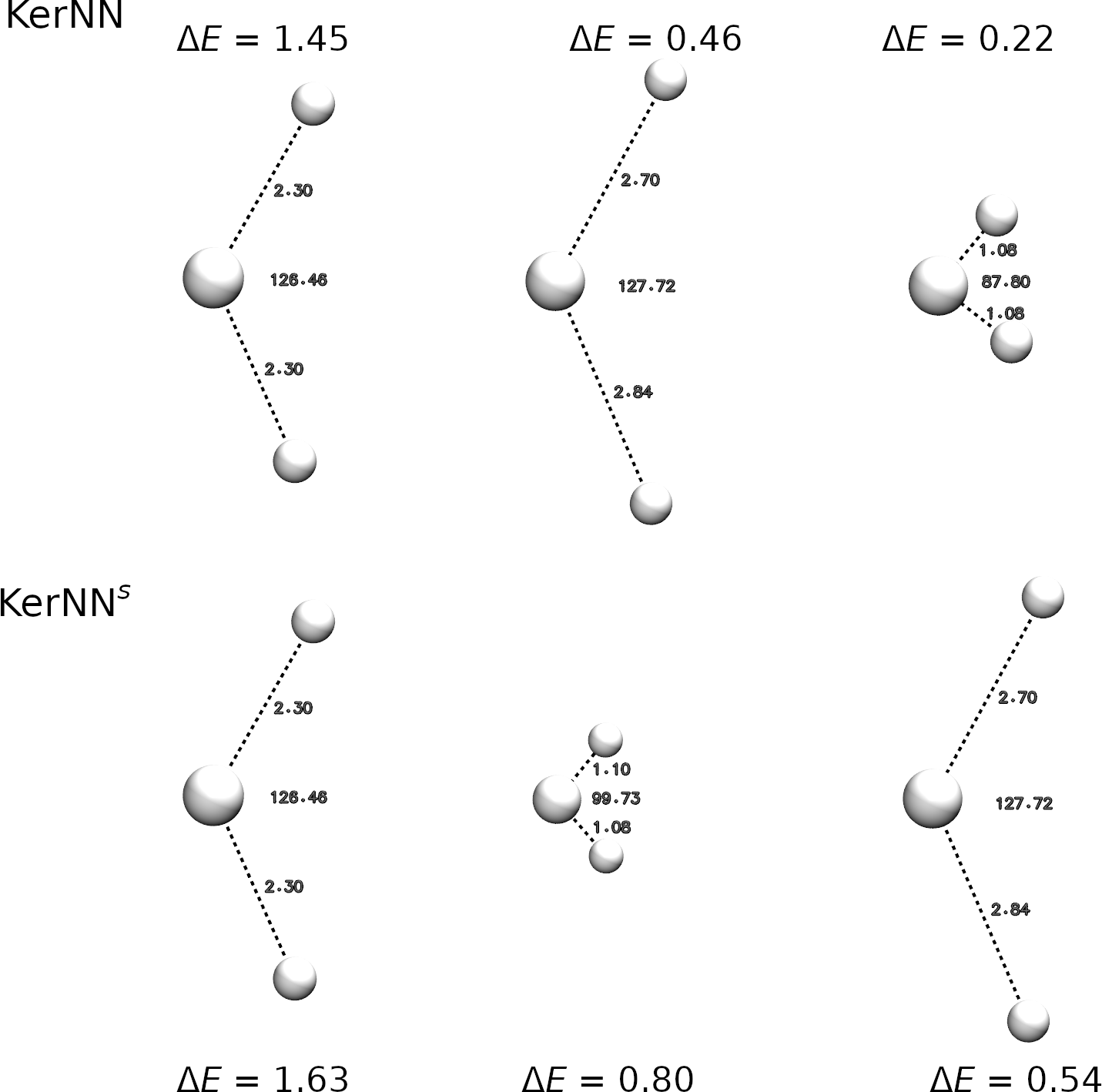}
\caption{Test set structures with the highest prediction error as compared to direct UCCSD(T) \textit{ab initio} calculations for both KerNN$^{\rm ns}$ and KerNN$^{\rm s}$. For each structure the associated bond length (in \AA), the H-He-H angle (in $^\circ$) and the absolute energy difference between reference and predicted energy (in kcal/mol) are shown. Two out of three outliers are the same in KerNN$^{\rm ns}$ and KerNN$^{\rm s}$. }
\label{sifig:outliers_heh2p}
\end{figure}

\begin{figure}[h]
\centering
\includegraphics[width=1.0\textwidth]{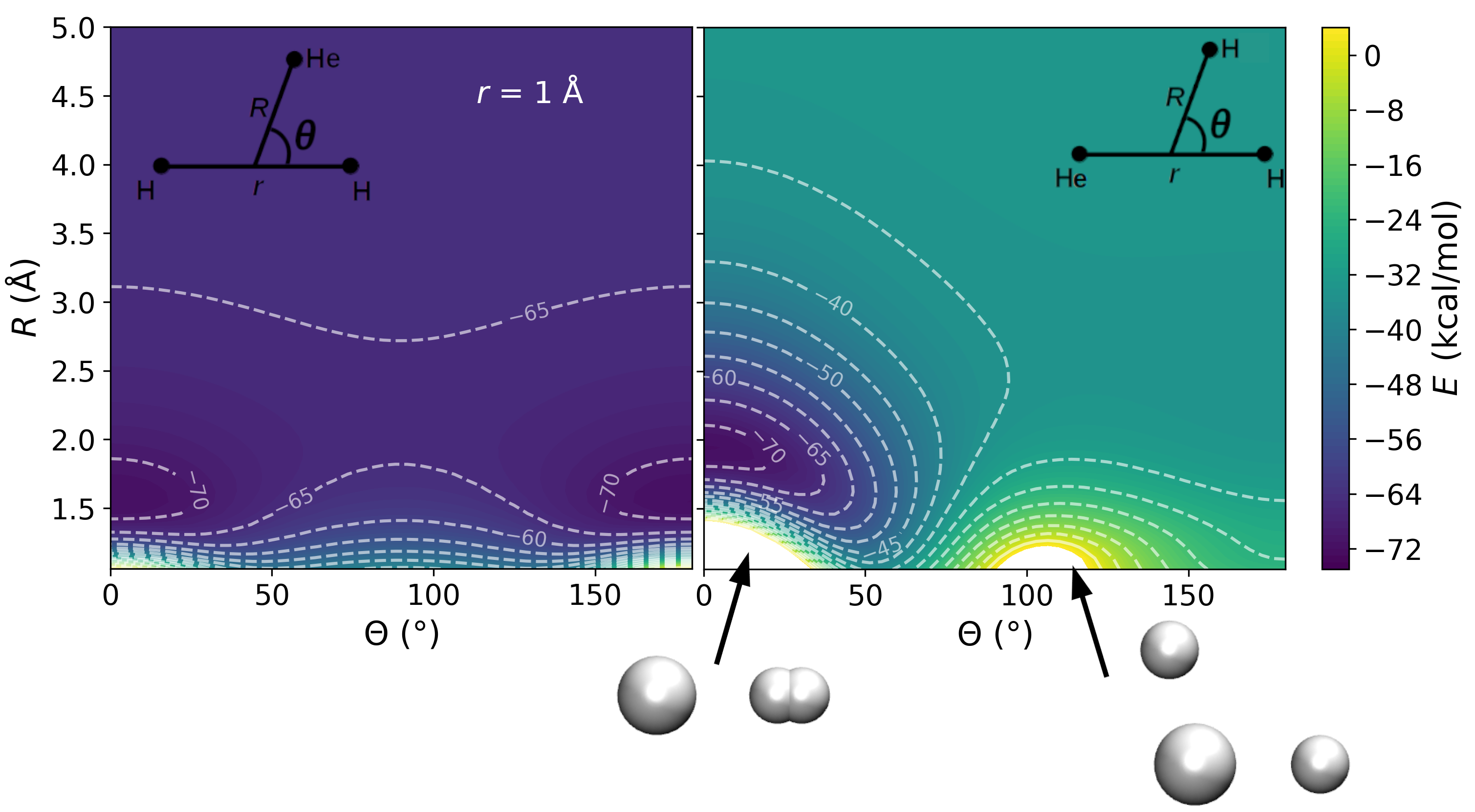}
\caption{Two-dimensional cuts through the 3D KerNN$^{\rm ns}$ PES for He + H$_2^+$ (left) and HeH$^+$ + H (right) channels. The separation of the diatomics, H$_2^+$ and HeH$^+$ is fixed at $\sim 1$\AA. Repulsive regions are illustrated with a schematic and energies that exceed 5~kcal/mol are shown in white.}
\label{sifig:kernn_2dpes_heh2+}
\end{figure}

\begin{figure}[ht]
\centering
\includegraphics[width=1.0\textwidth]{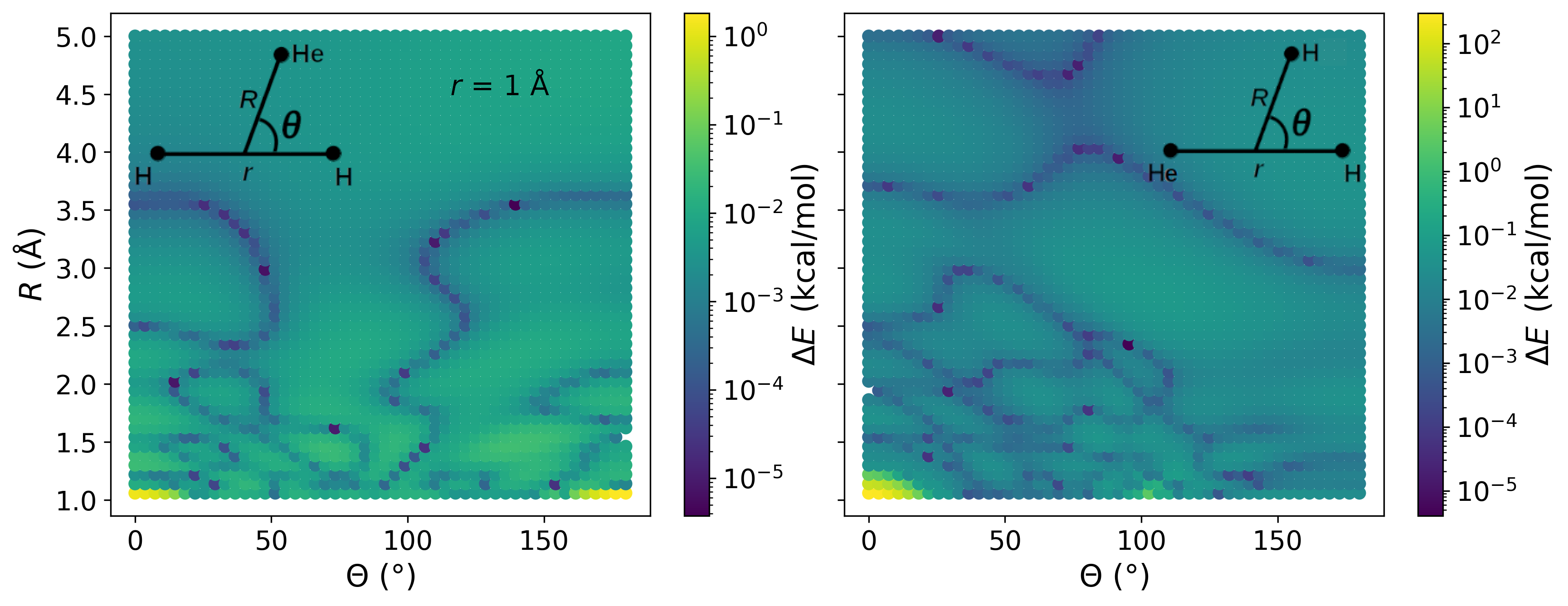}
\caption{Absolute errors of the KerNN$^{\rm ns}$ PES for He + H$_2^+$ (left) and HeH$^+$ + H (right) channels with respect to CCSD(T)/aug-cc-pV5Z reference energies for 2500 grid points. The largest differences are found in the repulsive regions. The predictions yield MAE($E$) of 0.05 and 0.70~kcal/mol for the He + H$_2^+$ and HeH$^+$ + H channels, respectively. Excluding the ten predictions with largest error for the latter (residing exclusively in the repulsive region) yields a MAE($E$) of 0.06~kcal/mol.}
\label{sifig:kernn_2dpes_heh2+_diff}
\end{figure}

\begin{figure}[h]
\centering
\includegraphics[width=1.0\textwidth]{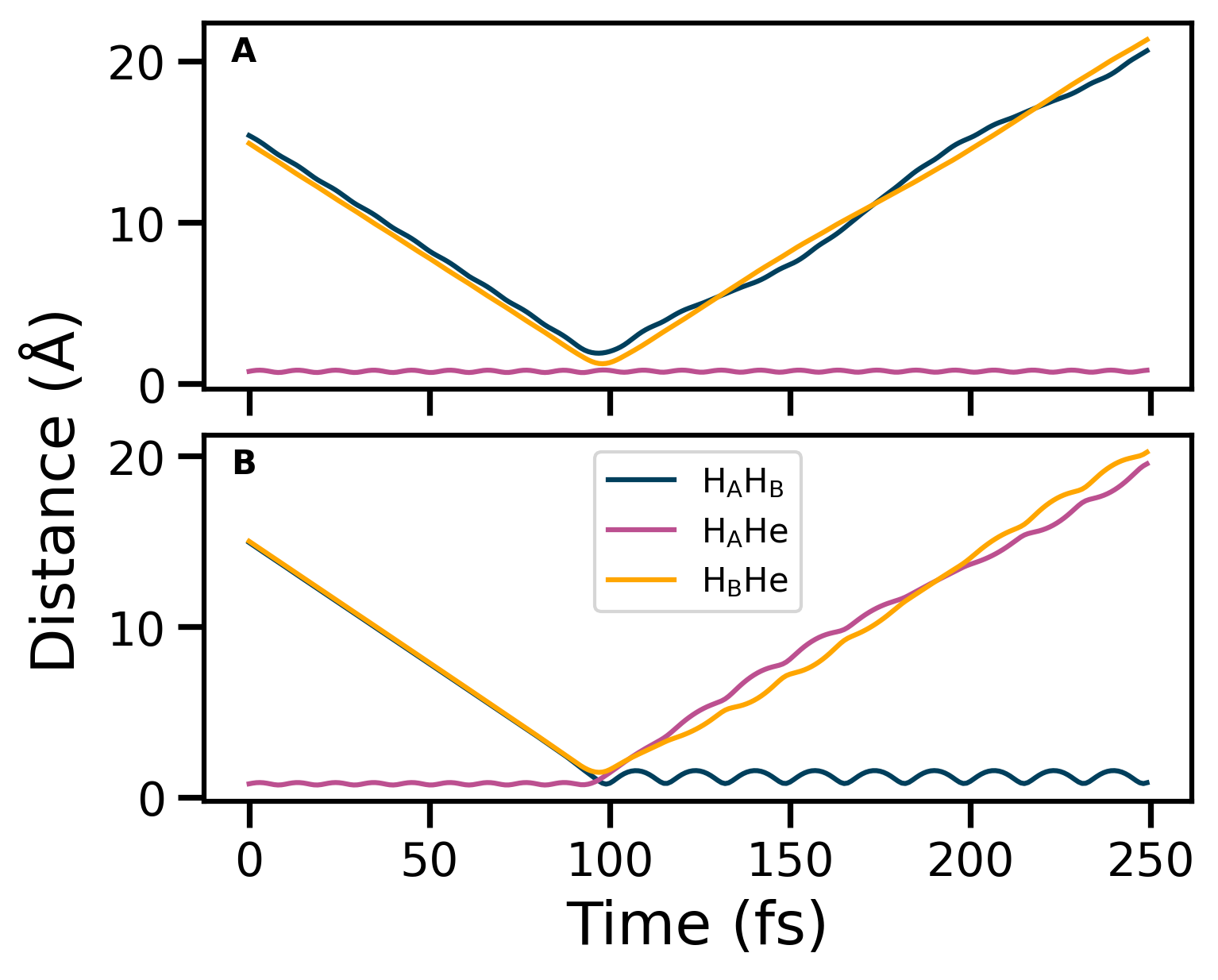}
\caption{Exemplary non-reactive (A) and reactive (B) collision trajectories obtained from KerNN$^{\rm ns}$. Initially, H$_{\rm A}$He forms the bonded diatom with a vibrational energy that has been assigned by drawing random momenta from a Maxwell-Boltzmann distribution corresponding to 300~K (translation and rotation have been projected out).  H$_{\rm B}$, initially placed at a distance of 15~\AA\@ from the CoM of the diatom, is accelerated towards the CoM of the diatom with a kinetic energy corresponding to $\sim 1$~eV.}
\label{sifig:reac_traj_heh2+}
\end{figure}

\begin{longtable}[c]{llllrr}\caption{Bound energy levels from DVR3D calculations for
e- and f-parity, \textit{ortho} (o) and \textit{para} (p) He--H$_2^+$ for $J=0$ and $J=1$  in cm$^{-1}$. The results from the KerNN$^{\rm ns}$ PES trained on UCCSD(T)/aug-cc-pV5Z level of theory data are compared to theoretical results derived from a FCI/aug-cc-pV5Z PES.\cite{MM.heh2:2019}}\\\toprule
o/p &   $J$ &  parity & $n$  & FCI/DVR3D     & KerNN$^{\rm ns}$/DVR3D  \\\midrule
o	&	0	&	e	&	1	&	-1793.7632	&	-1793.3578	\\
o	&	0	&	e	&	2	&	-1062.6727	&	-1064.2547	\\
o	&	0	&	e	&	3	&	-657.6932	&	-658.5876	\\
o	&	0	&	e	&	4	&	-538.1281	&	-541.3037	\\
o	&	0	&	e	&	5	&	-254.3251	&	-257.9835	\\
o	&	0	&	e	&	6	&	-190.0415	&	-193.8107	\\
o	&	0	&	e	&	7	&	-48.5279	&	-54.2574	\\
o	&	0	&	e	&	8	&	19.0868	&	15.2201	\\
o	&	0	&	e	&	9	&	40.0360	&	35.9073	\\
o	&	0	&	e	&	10	&	55.8602	&	50.0858	\\\midrule
o	&	1	&	e	&	1	&	-1785.5973	&	-1783.5868	\\
o	&	1	&	e	&	2	&	-1153.7813	&	-1152.7036	\\
o	&	1	&	e	&	3	&	-1055.3035	&	-1055.2840	\\
o	&	1	&	e	&	4	&	-650.0284	&	-649.3253	\\
o	&	1	&	e	&	5	&	-564.6730	&	-565.8948	\\
o	&	1	&	e	&	6	&	-531.0451	&	-532.6107	\\
o	&	1	&	e	&	7	&	-263.2100	&	-264.2067	\\
o	&	1	&	e	&	8	&	-248.8408	&	-250.9070	\\
o	&	1	&	e	&	9	&	-186.1513	&	-188.1117	\\
o	&	1	&	e	&	10	&	-174.2152	&	-175.6023	\\
o	&	1	&	e	&	11	&	-68.9712	&	-71.8441	\\
o	&	1	&	e	&	12	&	-45.1411	&	-49.2928	\\
o	&	1	&	e	&	13	&	-27.0299	&	-29.0847	\\
o	&	1	&	e	&	14	&	22.8241	&	20.0109	\\
o	&	1	&	e	&	15	&	30.5164	&	26.4763	\\
o	&	1	&	e	&	16	&	43.4379	&	40.8763	\\
o	&	1	&	e	&	17	&	54.5753	&	50.1988	\\
o	&	1	&	e	&	18	&	57.2317	&	52.8676	\\
o	&	1	&	f	&	1	&	-1153.4648	&	-1152.3852	\\
o	&	1	&	f	&	2	&	-563.7240	&	-564.9356	\\
o	&	1	&	f	&	3	&	-262.8397	&	-263.8038	\\
o	&	1	&	f	&	4	&	-175.8616	&	-177.0772	\\
o	&	1	&	f	&	5	&	-68.7581	&	-71.6364	\\
o	&	1	&	f	&	6	&	-27.5094	&	-29.5545	\\
o	&	1	&	f	&	7	&	29.9118	&	25.7797	\\
o	&	1	&	f	&	8	&	54.6476	&	50.1420	\\\midrule
p	&	0	&	e	&	1	&	-1793.7639	&	-1793.3585	\\
p	&	0	&	e	&	2	&	-1062.7083	&	-1064.2907	\\
p	&	0	&	e	&	3	&	-659.6723	&	-660.5809	\\
p	&	0	&	e	&	4	&	-538.8389	&	-542.0167	\\
p	&	0	&	e	&	5	&	-300.6517	&	-303.4586	\\
p	&	0	&	e	&	6	&	-214.3129	&	-218.8744	\\
p	&	0	&	e	&	7	&	-109.8684	&	-114.4646	\\
p	&	0	&	e	&	8	&	-60.7249	&	-65.7386	\\
p	&	0	&	e	&	9	&	-16.0670	&	-21.9619	\\
p	&	0	&	e	&	10	&	-1.1504	&	-7.3349	\\\midrule
p	&	1	&	e	&	1	&	-1785.5980	&	-1783.5875	\\
p	&	1	&	e	&	2	&	-1153.7480	&	-1152.6700	\\
p	&	1	&	e	&	3	&	-1055.3364	&	-1055.3174	\\
p	&	1	&	e	&	4	&	-651.8816	&	-651.1921	\\
p	&	1	&	e	&	5	&	-562.5239	&	-563.7546	\\
p	&	1	&	e	&	6	&	-531.5656	&	-533.1320	\\
p	&	1	&	e	&	7	&	-295.1356	&	-296.3543	\\
p	&	1	&	e	&	8	&	-212.6546	&	-213.8757	\\
p	&	1	&	e	&	9	&	-209.9754	&	-212.8800	\\
p	&	1	&	e	&	10	&	-154.5125	&	-155.3351	\\
p	&	1	&	e	&	11	&	-105.5427	&	-108.6013	\\
p	&	1	&	e	&	12	&	-57.0589	&	-60.4399	\\
p	&	1	&	e	&	13	&	-14.3679	&	-18.6642	\\
p	&	1	&	e	&	14	&	-0.4477	&	-5.0399	\\
p	&	1	&	f	&	1	&	-1153.4283	&	-1152.3484	\\
p	&	1	&	f	&	2	&	-561.2379	&	-562.4576	\\
p	&	1	&	f	&	3	&	-212.4739	&	-213.6612	\\
p	&	1	&	f	&	4	&	-154.6979	&	-155.4914	\\\bottomrule
\label{sitab:heh2+_boundstates}
\end{longtable}

\begin{figure}[h]
\centering
\includegraphics[width=0.95\textwidth]{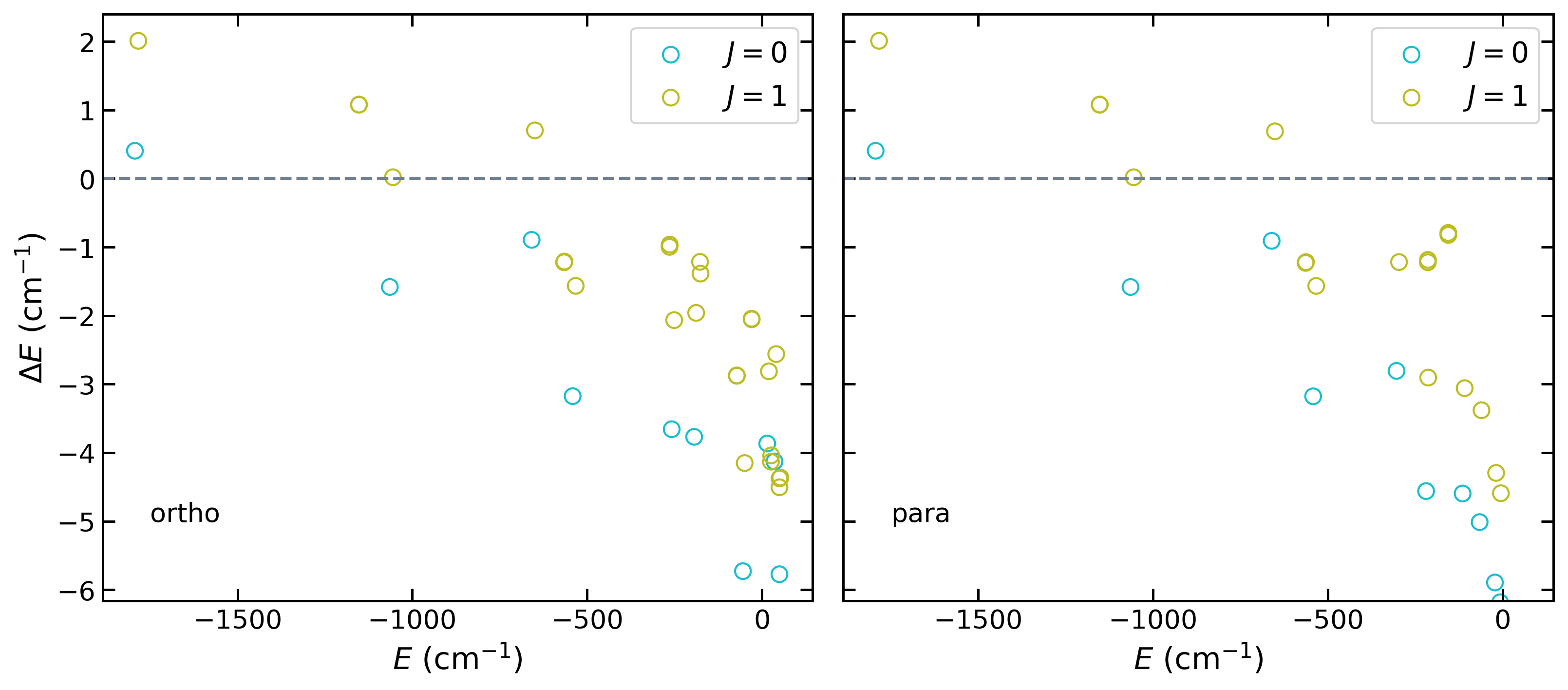}
\caption{Comparison of the bound energy levels from DVR3D calculations
  for $J=0$ and 1, \textit{e-} and \textit{f-} parity, and
  \textit{ortho-} and \textit{para-}HeH$_2^+$ as determined on the
  KerNN$^{\rm ns}$(UCCSD(T)) and a RKHS(FCI)
  PES\cite{MM.heh2:2019}. $\Delta E = E_{\rm KerNN} - E_{\rm RKHS}$
  and illustrates that KerNN$^{\rm ns}$ trained on UCCSD(T) typically
  underestimates the bound state energy in comparison to the FCI PES,
  in particular in the near-dissociative
  region.}\label{sifig:kernn_vs_rkhs_boundstates}
\end{figure}
\clearpage
\subsection*{Hydrogen Oxalate}
\begin{table}[h]
\centering
\begin{tabular}{rrrrrrr}
  \toprule
   & MAE($E$) & RMSE($E$) & MAE($F$) & RMSE($F$) & MAE($\mu$) & RMSE($\mu$) \\
  \midrule
\textbf{KerNN$^{\rm ns}$}   & 0.013 & 0.033 & 0.089 & 0.187 & 0.0014 & 0.0027\\
\textbf{PhysNet} & 0.009 & 0.047 & 0.065 & 0.459 & 0.0018 & 0.0054 \\
\bottomrule  
\end{tabular}
\caption{Averaged test set errors for the KerNN$^{\rm ns}$ and PhysNet
  PES trained on the MP2/aug-cc-pVTZ level data set. Energy, force and
  dipole moment errors are given in kcal/mol, kcal/mol/\AA, and Debye,
  respectively.}
\label{sitab:oxa1}
\end{table}
\clearpage
\bibliography{references}

%TC:endignore
\end{document}